\documentclass[journal,10pt]{IEEEtran}
\usepackage{epsfig}
\usepackage{graphicx}
\usepackage{orcidlink}
\usepackage{subfigure}
\usepackage{amssymb}
\usepackage{makeidx}
\usepackage{amsmath}
\usepackage{amsthm}

\newtheorem{theorem}{Theorem}

\newtheorem{lemma}{Lemma}

\usepackage{amssymb}
\usepackage{makeidx}
\usepackage{amsmath}
\usepackage{graphicx}
\usepackage{epsf}
\usepackage{epsfig}
\usepackage{psfig}
\usepackage{ccaption}
\usepackage{array}
\usepackage{tabularx}
\usepackage{multirow}
\usepackage{epsfig}
\usepackage{cite}
\usepackage{procedure}
\usepackage{algorithm}
\usepackage{algpseudocode}

\usepackage{geometry}

\newgeometry{left=1.4cm,right=1.4cm,bottom=1.8cm,top=1.2cm}
\begin{document}

\title{Wideband Integrated Sensing and Communications: Spectral Efficiency and Signaling Design}
\author{
Henglin Pu,
Zhu Han, \IEEEmembership{Fellow, IEEE},
Athina P. Petropulu, \IEEEmembership{Fellow, IEEE},
and Husheng Li, \IEEEmembership{Senior Member, IEEE}

\thanks{Henglin Pu and Husheng Li are with the Elmore Family School of Electrical and Computer Engineering, Purdue University, West Lafayette, Indiana, USA (email:
 pu36@purdue.edu, husheng@purdue.edu).}%
\and
\thanks{Zhu Han is with the Department of Electrical Engineering, University of Houston (email: zhan2@uh.edu).} \and
\thanks{Athina P. Petropulu is with the Department of Electrical and Computer Engineering, Rutgers The State University of New Jersey, New Brunswick, NJ, USA
(email: athinap@soe.rutgers.edu). }
\and
\thanks{Husheng Li is with the School of Aeronautics and
 Astronautics, Purdue University, West Lafayette, Indiana, USA (email: husheng@purdue.edu).}%
}
\maketitle

\begin{abstract}
In integrated sensing and communications (ISAC), a distinguishing feature of 6G wireless networks, the main challenge lies in integrating the two distinct functions of sensing and communication within the same waveform. In this paper, the ISAC waveform synthesis is studied in the wideband regime, since a large bandwidth can simplify the analysis and is justified by the employment of millimeter wave or higher frequency band. Standard orthogonal frequency division multiplexing (OFDM) signaling is assumed, and the wideband analysis of sensing is a counterpart of the existing studies on wideband communications. It is proposed that the phase over such OFDM subcarriers is for modulating communication messages while the power spectral density (PSD) is shaped for the sensing performance. Beyond OFDM, we further reveal a duality between the proposed PSD-shaping rule and the orthogonal time frequency space (OTFS) waveform. Flattening the OTFS delay-axis PSD produces the same integrated sidelobe level (ISL) reduction effect in the delay–Doppler domain as PSD control achieves for OFDM in the frequency domain. To balance communication and sensing performance over frequency-selective channels, we propose a low-complexity, water-filling-like allocator with an explicit PSD-flatness (variance) constraint. The performance of the proposed wideband ISAC scheme is demonstrated using both numerical simulations and hardware experiments. 
\end{abstract}


\thispagestyle{empty}

\section{Introduction}

% Integrated sensing and communications (ISAC) is a technology that enables both functions in the same waveform: the forward propagation of modulated electromagnetic (EM) wave brings communications messages to destinations, while the backward propagation of reflected EM waves (if there is any) fetches the environmental information to the transmitter or other sensing receivers for sensing.

Integrated sensing and communications (ISAC) is a technology that enables both functions of sensing and communication in the same waveform. During forward propagation, the modulated electromagnetic (EM) wave brings communications messages to destinations. Portions of the wave that reflect off surrounding objects travel back toward the transmitter or other sensing receivers, where they are processed to extract information about the environment. ISAC is expected to be a featuring technology in 6G communication networks, due to its significant spectral, power and hardware efficiencies~\cite{ISAC1}. 

The major challenge in ISAC is how to integrate both functions of communications and sensing in the same waveform. The optimal signaling of either communications or sensing is typically complicated. For example, the theoretically optimal waveform of communications is Gaussian random codes that are difficult to implement in practical systems, subject to additive white Gaussian noise (AWGN) channel. On the other hand, there has not been a well recognized optimal radar sensing waveform partially due to the diversity of radar design metrics. Hence, it is difficult to synthesize the waveform that achieves the optimal trade-off between communications and sensing, for generic situations. 

To fill this research gap in this paper, we will study the waveform synthesis of ISAC in the wideband regime. In current communication system, a total of 400MHz bandwidth can be used for spectrum access of each service provider~\cite{3gpp-38101-2-R19}. Radar sensing system may use much more (e.g., 4GHz bandwidth in the TI 77GHz millimeter wave radar set). Such a wide frequency band provides plenty of degrees of freedom (DoFs), each of which has very low available signal power. In this wideband regime, the optimal signaling can be substantially simplified. For example, it has been shown that, when the bandwidth tends to infinity, the antipodal signals, such as binary phase shift keying (BPSK) for real AWGN channels, can achieve the maximum channel capacity for communications \cite{Golay1949,Turin1959}. When the spectral efficiency is taken into account with a large but finite bandwidth, it is shown in \cite{Verdu2002} that in complex AWGN channels the quadrature phase shift keying (QPSK) achieves the maximal spectral efficiency in terms of the increasing slope, thus minimizing the demand of frequency band. The signaling of phase modulation with constant modulus brings good news to the design of wideband communications. A detailed analysis on the spectral efficiency of communications in the wideband regime (such as the minimum energy per bit and the increasing slope of spectral efficiency) has been carried out in \cite{Verdu2002}, from which we can leverage many existing conclusions. 

However, there is no such a counterpart study for wideband radar sensing -- for example, on the minimum energy required to achieve a certain ranging precision. As the bandwidth increases to infinity, does the sensing performance become asymptotically perfect, or is it fundamentally limited by bounded energy? In this paper, we will also study the radar sensing performance and signaling in the wideband regime. For the generic case, the sensing performance is mostly dependent on the power spectral density (PSD). The evidences include the following performance metrics:
\begin{itemize}
\item {\em Ranging resolution:} When the radar pulses are close to rectangular, the ranging accuracy characterized by the root mean square error is $\delta T_R=\frac{c}{\beta \sqrt{\frac{2E_p}{N_0}}}$ (Eq. (6.9) in \cite{Skolnik2001}),
where $c$ is the light speed, $E_p$ is the pulse energy, $N_0$ is the noise power spectral density, and $\beta$ the effective bandwidth defined as
$\beta=\frac{1}{E}\int_{-\infty}^{\infty}(2\pi f)^2|S(f)|^2df,$
where $S(f)$ is the PSD. We observe that the phase of the radar waveform does not affect the ranging performance. Only the PSD plays a role.

\item {\em Sidelobe mitigation:} The sidelobes of the autocorrelation function of radar waveform are of critical importance as a design metric, since a sidelobe may be confused with a weak target. It is well known that the time-domain autocorrelation and PSD form a Fourier transformation pair. Therefore, one can refine the design of PSD (e.g., using a proper window function) in order to better mitigate the sidelobes.
\end{itemize}

Summarizing the above evidences, we can consider a wideband ISAC scheme in which the communication modulation is over the phase in the frequency domain (similarly to the orthogonal frequency division multiplexing (OFDM)) while the power spectrum is optimized for radar sensing, thus decoupling the designs for communications and sensing in ISAC and facilitating their coexistence. In this paper, we will substantiate this idea and demonstrate the proposed scheme using numerical simulations, answering the question: \textit{Is OFDM+QPSK simple but good design for wideband ISAC?} We further show that the flat-PSD rule that minimizes the integrated sidelobe level (ISL) in OFDM has a direct analogue in orthogonal time frequency space (OTFS)~\cite{OTFS3}, a waveform that has attracted significant recent interest and is well suited to high-mobility communications. Specifically, enforcing a flat PSD along the delay axis in the delay–Doppler domain yields low sidelobes, mirroring the effect of PSD control for OFDM in the frequency domain. To coordinate sensing and communication under frequency-selective channels, we also propose a low-complexity, water-filling-like allocator with an explicit PSD-flatness control that shapes the ambiguity function, yielding lower sidelobes. Simply put, this paper makes the following contributions:

\begin{itemize}
\item We propose an OFDM-based wideband ISAC design that embeds information in subcarrier phases while shaping the transmit PSD for sensing, cleanly decoupling communication modulation from sensing optimization. This separation yields constant-modulus signaling for robust hardware operation and enables principled PSD control to suppress range sidelobes.

\item We develop a wideband sensing analysis that links delay-estimation accuracy and sidelobe behavior to the PSD, clarifying when and why PSD flatness improves autocorrelation properties. Beyond OFDM, we show that flattening the OTFS delay-axis PSD achieves the same ISL reduction in the delay–Doppler domain as PSD shaping does in the frequency domain, yielding a unified low-sidelobe design rule across both modulations.

\item  We introduce a water-filling-like power allocator with an explicit PSD-flatness (variance) constraint to balance rate and sensing. The method operates with near–water-filling complexity, and admits simple parameter tuning.

\item Numerical studies and a mmWave SDR prototype show consistent reductions in integrated/peak sidelobe levels and a narrower ambiguity mainlobe. The hardware results corroborate the wideband analysis, confirming practical sensing gains from PSD shaping.
\end{itemize}

The remainder of the paper is organized as follows. The related works are introduced in Section \ref{sec:related}. The system model is given in Section \ref{sec:system}. Then, the wideband sensing and wideband ISAC are discussed in Sections \ref{sec:wide_sensing} and \ref{sec:wideband}, respectively. Finally, the numerical simulations and conclusions are given in Sections \ref{sec:numerical} and \ref{sec:conclusions}, respectively.

\begin{figure}[!t]
\centering
\includegraphics[width=0.36\textwidth]{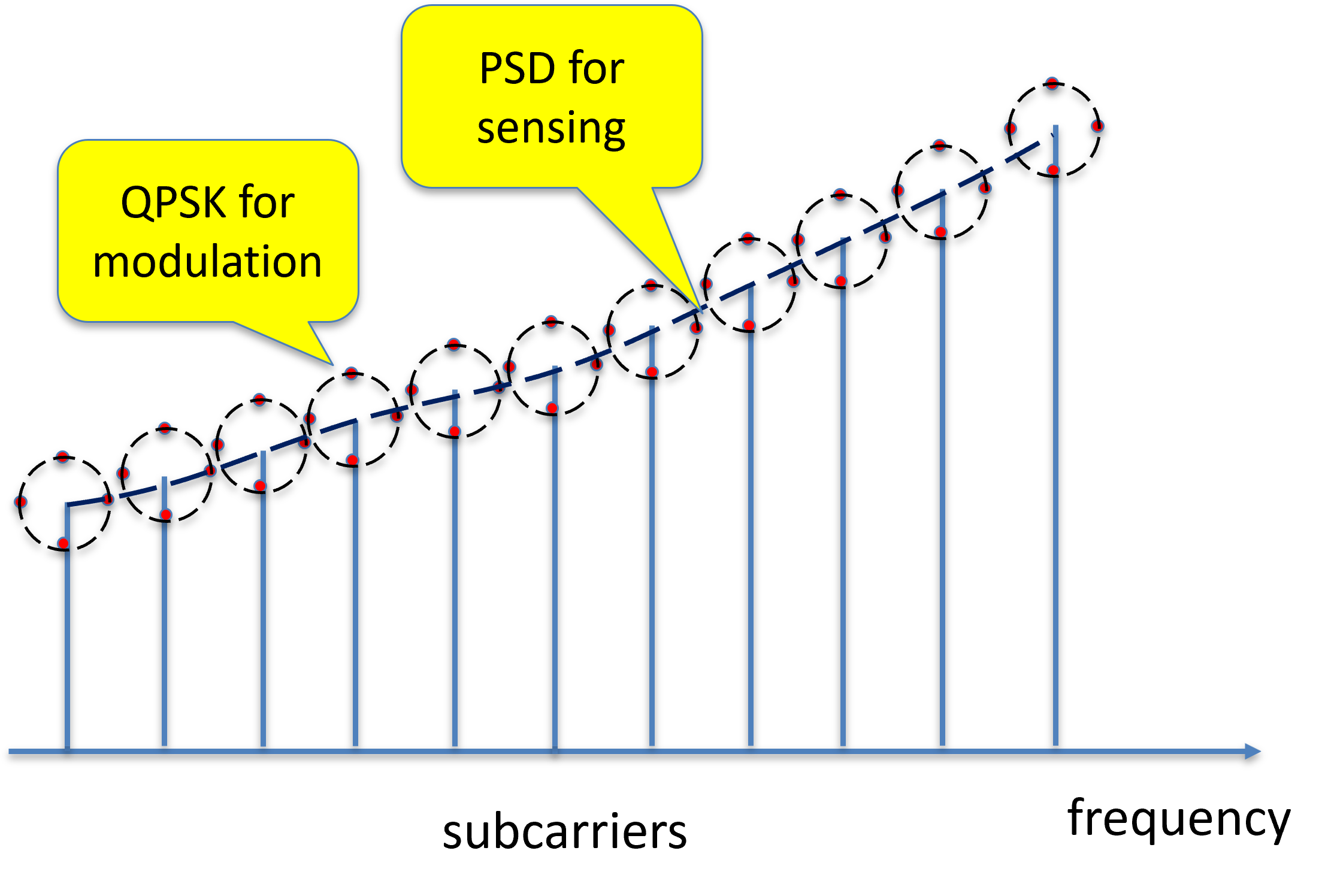}
\caption{QPSK modulation + PSD shaping for ISAC}
\label{fig:ISAC}
\end{figure}

\section{Related Works}\label{sec:related}
Comprehensive surveys on ISAC can be found in \cite{Zheng2019,FanLiu2020,Ma2020,FanLiu2023, ISAC_book1, ISAC_book2, ISAC_book3}. Most studies on the waveform design ISAC are either sensing-centric or communication-centric. In sensing-centric designs, a communication messages are embedded in the traditional radar waveforms; e.g., modulation over the chirp rate of frequency modulation continuous wave (FMCW) radar \cite{HLi_Globecom_2019_radar} or index modulation \cite{Temiz2023}. A more advanced approach is to use watermark coding to hide communication messages in the radar waveform, with a minimum performance degradation of radar sensing \cite{LiWatermark2023}. Meanwhile, the approach of symbol-level precoding (SLP) \cite{AngLi2020} based ISAC in \cite{RangLiu2022,ZihanLiao2023} is similar to the watermarking approach, while it does not consider the distortion on the radar waveform. In communication-centric ISAC, the communication waveform could be directly used for sensing (e.g., the Wi-Fi sensing of gestures \cite{JiangWiFi2018} and cellular-signal-based positioning \cite{Peral2017}). Far fewer studies have been dedicated to first-principle-based ISAC waveform designs, in which communications and sensing are flexibly balanced. 

Suppressing sensing side-lobes in communication-centric ISAC systems is particularly challenging when the transmitter carries random data streams \cite{tradeoff1,tradeoff2,tradeoff3}. This difficulty epitomizes the fundamental trade-off between communication and sensing performance, itself a manifestation of the broader deterministic-versus-random transmission dilemma \cite{tradeoff4}. Higher-order QAM constellations raise spectral efficiency, yet their non-constant-modulus symbols enlarge side-lobes and degrade sensing accuracy \cite{tradeoff1}. Conversely, constant-modulus QPSK preserves low side-lobes and sharp sensing resolution at the cost of a lower data rate. Recent studies tackle this trade off by optimizing the transmit signal in OFDM ISAC systems \cite{tradeoff1,tradeoff2,tradeoff5,tradeoff7}. \cite{tradeoff4,tradeoff5,tradeoff6} formulate ISAC trade-off optimization problems that reshape the input data distribution to balance the tradeoff. \cite{tradeoff2} designs a symbol level precoder that minimizes the range Doppler integrated side lobe level while meeting multiuser QoS targets. \cite{tradeoff1} applies probabilistic constellation shaping to maximize rate under radar AF variance limits, and \cite{tradeoff7} uses trellis shaping to reduce integrated side lobes. However, most of these work formulate the problems as nonconvex or NP hard, especially under strict radar metrics, which makes them computationally intensive. 

This prompts a key question: can we craft a simpler yet effective solution for wideband ISAC? The broadcast channel framework is a powerful tool for designing dedicated ISAC waveforms~\cite{LiBroadcast2023_1,LiBroadcast2023_2}. 

Studies on wideband communications mainly stems from the employment of ultra-wideband (UWB) signaling \cite{Sahinoglu2011}. The information theoretic aspect of wideband signaling has been comprehensively studied in \cite{Verdu2002}. An important observation is that a simple phase modulation is asymptotically optimal in terms of spectral efficiency in the wideband regime \cite{Golay1949,Turin1959}. A preliminary version of this study was presented at a conference in \cite{conf_ver}, where the baseline OFDM+QPSK design was introduced and validated through numerical simulations.

\section{System Model}\label{sec:system}

In our system, we adopt a wideband ISAC scheme in which communication data are conveyed via frequency-domain phase modulation (akin to OFDM), while the transmit power spectrum is optimized for radar sensing, thereby decoupling the two designs and facilitating their coexistence, as illustrated in Fig.~\ref{fig:ISAC}.

We adopt a quasi-static per-pulse model (point scatterers, time-invariant within a pulse), allowing us to omit explicit time dependence and state per-pulse relations generically. The dynamic case follows by introducing per-path Doppler across slow time while leaving the per-pulse baseband unchanged; details are omitted for brevity.
 We consider a continuous waveform, $x(t)$, whose support is $[0,T_p]$, where $T_p$ is the pulse (symbol) duration. The available bandwidth is $W$. We consider the following two tasks of radar sensing:
\begin{itemize}
\item {\em Ranging:} For the task of ranging, the target is considered as a point. The received signal, reflected by the target, is given by 
\begin{eqnarray}\label{eq:received_reflection_signal}
y(t)=Ax(t-\tau)+n(t),
\end{eqnarray}
where $\tau$ is the round trip time of the signal, $A$ is the signal attenuation, and $n$ is the thermal noise at the sensing receiver. The goal is to estimate time delay $\tau$ (and thus the distance to the target).

\item {\em Imaging:} When the target is large and within a region $\Omega$, the receiver needs to be spatially dispersed (e.g., an antenna array). Therefore, the received signal at position $\mathbf{r}$ and time $t$ is given by 
\begin{eqnarray}\label{eq:imaging}
y(\mathbf{r},t)=A\int_{\Omega}J(\mathbf{r}')x(t-\tau(\mathbf{r},\mathbf{r}'))d\mathbf{r}'+n(\mathbf{r},t),
\end{eqnarray}
where $A$ is the signal antennuation due to the path loss, $J(\mathbf{r}')$ is the reflection coefficient, $\tau(\mathbf{r}')$ is the signal delay when reflected at position $\mathbf{r}'$ and $n$ is the thermal noise depending on position and time. The goal is to estimate $|J(\mathbf{r}')|$, namely the magnitude of the reflection coefficient. 
\end{itemize}

Another important task of radar sensing is the speed estimation, which is based on the estimation of Doppler shift. It is well known that the performance of Doppler shift estimation is determined by the time duration of measurement (the slow-time data). For example, when a rectangular pulse is used, the root mean square error of Doppler frequency shift estimation is given by $\delta f=\frac{1}{\alpha\sqrt{\frac{2E}{N_0}}}$ ((6.22) in \cite{Skolnik2001}),
where $E$ is the received signal energy and $\alpha=\frac{\int_{-\infty}^\infty (2\pi t)^2 s^2(t)dt}{\int_{-\infty}^\infty s^2(t)dt}$ is called the effective time duration of the time-domain signal $s(t)$. Therefore, it is largely independent of the signal bandwidth and thus is not included in the discussion of this paper.

\subsection{Input Output Relation of OFDM}

Consider a standard OFDM modulation with symbol duration \( T_o = T_{\text{cp}} + T \), where \( T_{\text{cp}} \) and \( T \) denote the CP and data symbols duration, respectively. We assume that \( X_{mn} \) denotes the data symbol of the \( m \)-th subchannel in the \( n \)-th time interval. The continuous-time OFDM transmitted signal with CP is given by
\begin{equation}
s(t) = \sum_{n=0}^{N-1} \sum_{m=0}^{M-1} x_{n,m} \, \text{rect}(t - nT_o) e^{j2\pi m \Delta f (t - T_{\text{cp}} - nT_o)},
\end{equation}
where \( \text{rect}(t) \) is one for \( t \in [0, T_o] \) and zero otherwise. We consider the channel as a \( P \)-tap time-frequency selective channel given by
\begin{equation}
h(t, \tau) = \sum_{i=0}^{P-1} h_i \delta(\tau - \tau_i) e^{j 2 \pi \nu_i t},
\end{equation}
where \( h_i \) is the complex channel gain (in sensing, $h_i$ subsumes target reflectivity and propagation loss), \( \nu_i\) and \( \tau_i\) denote the Doppler shift and delay, respectively. The signal received after the time-frequency selective channel is given by
\begin{equation}
\begin{aligned}
   y(t) &= \int h(t, \tau) s(t - \tau) d\tau + w(t) \\ &= \sum_{i=0}^{P-1} h_i s(t - \tau_i) e^{j2\pi \nu_i t} + w(t), 
\end{aligned}
\end{equation}
where $w(t)$ represents the AWGN. On the receiving end, the received baseband waveform is matched-filtered and sampled at a rate of \( 1/T \). Removing the samples in guard intervals and performing discrete Fourier transform (DFT) on the sampled signal, we obtain
\begin{equation}
\mathbf{Y}_{mn} = \mathbf{H}_{mn} \mathbf{X}_{mn} + \mathbf{N}_{mn},
\end{equation}
where \( \mathbf{N}_{mn}\) is a zero-mean complex Gaussian random variable with variance $\sigma_n^2$. $\mathbf{H}_{mn}$ represents the channel matrix, which is given by
\begin{equation}
\mathbf{H}_{mn} = \sum_{i=1}^{P} h_i[n] \exp\left( i2\pi m \frac{\tau_i[n]}{T} \right).
\end{equation}

\subsection{Input Output Relation of OTFS}

The transmission data $\mathbf{X}^{\mathrm{DD}} \in \mathbb{C}^{M_{\tau}\times N_{\nu}}$ is embedded in the DD domain with an $M_{\tau}\times N_{\nu}$ grid. Each grid cell spans $T$ seconds in the time dimension and $\Delta_f$ Hertz in the Doppler shift, uniquely corresponding to a specific delay and Doppler pair. $M_{\tau}$ and $N_{\nu}$ denote the number of delay and Doppler bins respectively. Then by applying an Inverse Symplectic Finite Fourier Transform (ISFFT), the DD domain signals are transformed into the TF domain:
\begin{equation}\label{eq:ISFFT}
\begin{array}{l}
\mathbf{X}^{\mathrm{TF}} = \mathbf{F}_{M_{\tau}}\mathbf{X}^{\mathrm{DD}} \mathbf{F}^H_{N_{\nu}}, 
\end{array}
\end{equation}
where $\mathbf{X}^{\mathrm{TF}} \in \mathbb{C}^{M_{\tau}\times N_{\nu}}$ is the time-frequency domain representation and $\mathbf{F}_N$ is the unitary discrete Fourier transform (DFT) matrix of size $N\times N$. Subsequently, the Heisenberg transform is employed and a transmit pulse-shaping filter $g_{tx}(t)$ supported in $[0,T]$ is utilized to generate the transmission signal $s(t)$ in time domain:
\begin{equation}
\begin{array}{l}
s(t) = \frac{1}{\sqrt{M_{\tau}}} \sum\limits^{N_{\nu}-1}\limits_{n=0} \sum\limits^{M_{\tau}-1}\limits_{m=0} \mathbf{X}^{\mathrm{TF}}[m,n] g_{tx}(t-nT)\\
\qquad \qquad \qquad \qquad \qquad \qquad \qquad \quad e^{j2\pi m\Delta_f(t-nT)}.
\end{array}
\end{equation}

On the communication receiver side, the received signal $r(t)$ is multiple delayed copies of transmitted ones: 
\begin{equation}
\begin{array}{l}
r(t) = \int_{\nu} \int_{\tau} h(\tau, \nu) s(t-\tau) e^{j2\pi \nu(t-\tau)}d\tau d\nu + w(t)\\
\qquad = \sum\limits^{P}\limits_{i=1} h_i e^{j2\pi v_i t} s(t-\tau_i) + w(t),
\end{array}
\end{equation}
where $w(t)$ represents the additive white Gaussian noise, $P$ denotes the number of paths, $h(\tau, \nu)$ represents the channel impulse response:
 \begin{equation}\label{eq:channelH}
\begin{array}{l}
h(\tau ,v) = \sum\limits^{P}\limits_{i=1} h_i \delta(\tau - \tau_i) \delta(\nu - \nu_i),
\end{array}
\end{equation}
where $h_i$ is the complex gain associated with the $i$-th path, while $\delta (\cdot)$ is the Dirac delta function. The delay $\tau_i$ and Doppler shift $\nu_i$ of the $i$-th path are modeled as integer multiples of the delay resolution and fractional Doppler resolution, given by $\tau_i = \frac{l_i}{M_{\tau} \Delta f}$ and $\nu_i = \frac{k_i + \kappa_i}{N_{\nu} T}$, respectively. Here, $l_i$ and $k_i$ are integers representing the delay and Doppler taps, while $\kappa_i \in (-0.5, 0.5]$ denotes the fractional Doppler shift of the $i$-th path. This time domain signal $r(t)$ is sampled and transformed to the TF domain signal $\mathbf{Y}^{\mathrm{TF}} \in \mathbb{C}^{M\times N}$ by applying the Wigner transform: 
\begin{equation}
\begin{array}{l}
\mathbf{Y}^{\mathrm{TF}}[m,n] = \int g^*_{rx}(t-nT)r(t)e^{-j2\pi m\Delta_f (t-nT)}dt, 
\end{array}
\end{equation}
where $g_{rx}(t)$ is the receive pulse-shaping filter. Next, the SFFT is applied to transform $\mathbf{Y}^{\mathrm{TF}}$ back to the DD domain:
\begin{equation}\label{eq:SFFT}
\begin{array}{l}
\mathbf{Y}^{\mathrm{DD}} = \mathbf{F}^{H}_{M_{\tau}
} \mathbf{Y}^{\mathrm{TF}} \mathbf{F}_{N_{\nu}}.
\end{array}
\end{equation}

\section{Wideband Analysis on Sensing}\label{sec:wide_sensing}
In this section, we carry out the analysis on the sensing performance in ISAC in the wideband regime, for both tasks of ranging and imaging.

\subsection{Ranging}
We first consider the task of ranging, whose performance includes the mean square error (MSE) and the sidelobe level.

\subsubsection{Ranging MSE}
We study the performance of ranging in the frequency domain, in which the received signal for the $n$-th pulse in the spectrum is given by the Fourier transform of (\ref{eq:received_reflection_signal})
\begin{eqnarray}\label{eq:received_signal_freq}
Y_n(jw)=Ae^{-j\tau w}X_n(jw)+N_n(jw),
\end{eqnarray}
where the noise $N_n(jw)$ is white. We notice that the round-trip time $\tau$ is the slope of the phase frequency response. The problem of ranging is thus converted to the estimation of phase response $e^{-j\tau w}$ in the frequency domain.

A lower bound for the MSE of ranging is disclosed in the following theorem, whose proof is given in Appendix \ref{appdx:proof}:
\begin{theorem}\label{thm:MSE}
For the ranging problem with received signal in (\ref{eq:received_signal_freq}), the MSE satisfies the following Cram\'er-Rao bound:
\begin{eqnarray}\label{eq:opt_MSE}
MSE \geq \frac{N_0Wc^2}{A^2\int_{\omega_c-\frac{W}{2}}^{\omega_c+\frac{W}{2}} w^2|X(jw)|^2d\omega}.
\end{eqnarray}
\end{theorem}

We notice that, when $W\rightarrow\infty$, the Cram\'er-Rao bound vanishes. It can be implemented using very narrow pulses of very high instantaneous power, due to the large bandwidth, such that very precise timing can be achieved against noise. This is in a sharp contrast to communications, in which the capacity is bounded by the limited power. Therefore, it is always desirable to increase the bandwidth for sensing, since the sensing MSE decreases inversely proportionally to the bandwidth, while the performance gain of communications brought by more bandwidth is marginal when the bandwidth is large. Although the power allocated to each degree of freedom vanishes when $W\rightarrow\infty$ (or equivalently the total noise power becomes infinite while the signal power is limited), the MSE of ranging still vanishes. This is because the ranging problem can be interpreted as a frequency estimation for sinusoidal functions. Consider $N$ degrees of freedom, the signal received over each degree of freedom is given by
\begin{eqnarray}\label{eq:freq_est}
Y_l=Ae^{-j\tau l\delta w}X_l+N_l,\qquad l=1,...,N,
\end{eqnarray}
which can be considered as a sinusoidal function of the frequency $l$ and $\tau \delta w$ can be considered as the “frequency of frequency”. The Cram\'er-Rao bound of frequency estimation is given by \cite{StevenKay1993}
\begin{eqnarray}
Var[f]&\propto& \frac{12}{\gamma}\frac{1}{N(N^2-1)}\nonumber\\
&\approx&\frac{N_0}{PN^2},
\end{eqnarray}
where $\gamma$ is the per-sample SNR and $P$ is signal power. This means that the standard deviation of frequency estimation error is proportional to $\frac{1}{N}$, which converges to 0 as $N\rightarrow\infty$. 

% \begin{remark}\label{remark:amp_est}
% Note that the amplitude estimation (i.e., estimating $A$ in (\ref{eq:freq_est})) is different from the range estimation. Consider the sufficient statistics $\bar{Y}=\frac{1}{N}\sum_{l=1}^N Y_l$. Suppose that $X_I$ is the same in all degrees of freedom. The corresponding signal is $A^2|X_l|^2=\frac{A^2P}{N}$ and the noise power is $\frac{\sigma_n^2}{N}$, thus resulting an SNR of $\frac{A^2P}{\sigma_n^2}$, which is independent of $N$. Therefore, for signal amplitude (common to all frequencies) estimation, infinite bandwidth (unbounded degrees of freedom) does not nullify the estimation error, given a limited power. Therefore, we can claim that the amplitude estimation is power-limited.

% In a contrast, the range estimation leads to an asymptotically zero error as the bandwidth tends to infinity, even if the total power is limited. An intuitive explanation is that, in the frequency-domain representation, the delay is scaled by the frequency. Thus, the bandwidth plays the role of power in the amplitude estimation, and infinite bandwidth is equivalent to infinite power. Note that the range estimation error is bounded by neither bandwidth nor power. 
% \end{remark}

\subsection{Spectral Efficiency of Ranging}
In the context of communications, the concept of spectral efficiency is very natural, since it means the data rate over each degree of freedom. The performance metric, defined as the total data rate, is cumulative over all the degrees of freedom. However, in the context of sensing, the performance metric (e.g., the MSE of ranging) is not cumulative: in (\ref{eq:opt_MSE}), assume that the PSD $X(j\omega)$ is a constant $P_0$. Then, the denominator is
\begin{eqnarray}
\int_{\omega_c-\frac{W}{2}}^{\omega_c+\frac{W}{2}} w^2|X(jw)|^2d\omega&=& P_0\int_{\omega_c-\frac{W}{2}}^{\omega_c+\frac{W}{2}} w^2d\omega\nonumber\\
&=&P_0\left( \omega_c^2 W+\frac{1}{12} W^3\right).
\end{eqnarray}
Therefore, the lower bound for the ranging MSE equals 
\begin{eqnarray}
MSE&\geq& \frac{N_0 W c^2}{A^2P_0 \left(\omega_c^2 W+\frac{1}{12} W^3\right)}\nonumber\\
&=&\frac{N_0 c^2}{A^2P_0 \left(\omega_c^2 +\frac{1}{12} W^2\right)}=O\left(\frac{1}{W^2}\right),
\end{eqnarray}
by assume that $\frac{\omega_c}{W}$ remains to be a large number $\beta>0$ when $W\rightarrow\infty$. Hence, the performance lower bound for the standard deviation of ranging error is proportional to $\frac{1}{W}$. Since the Cramer-Rao bound is inversely proportional to the Fisher information, we can consider the Fisher information as being proportional to the square bandwidth $W^2$. Hence, we define the spectral efficiency $S_s$ for sensing as
\begin{eqnarray}\label{eq:efficiency_ranging}
S_s=\frac{\sqrt{I}}{W},
\end{eqnarray}
where $I$ is the Fisher information, which is assumed to be cumulative over the degrees of freedom in the frequency spectrum. The corresponding Fisher information $I$ for the parameter $\tau$ is therefore
\begin{equation}\label{eq:Fisher_info_appendix}
I
\;=\;
\frac{\,A^{2}\,P_{0}\,}{\,N_{0}\,W\,c^{2}\,}
\;
\Bigl(\,\omega_{c}^{2}+\tfrac{\,W^{2}\!}{\,12\,}\Bigr).
\end{equation}

Substituting \eqref{eq:Fisher_info_appendix} into \eqref{eq:efficiency_ranging}, the spectral efficiency for
sensing is given
\begin{align}\label{eq:Equation18_appendix}
S_{s}
&\;=\;
\frac{\sqrt{\,I\,}}{\,W\,}
\;=\;
\frac{\,A\,\sqrt{P_{0}}\,}{\,c\,\sqrt{N_{0}}\,W}\;
\sqrt{\frac{\omega_{c}^{2}\,W + \frac{W^{3}}{12}}
      {\,W\,}}
\\[5pt]
&=\frac{A\sqrt{P_0\left(\beta^2 +\frac{1}{12} \right)}}{c\sqrt{N_0} }.
\end{align}

\subsubsection{Sidelobe Level} An important concern of waveform is the sidelobes, since large sidelobes may be confused with weak targets, thus weakening the sensing capability. We consider the peak sidelobe level and assume random sequences. For a sequence $\mathbf{x}=(x_0,...,x_{N-1})$, the peak sidelobe is defined as
\begin{eqnarray}
    \mu_{\mathbf{x}}=\max_{0<u<N}\left|\sum_{j=0}^{N-u-1}x_jx_{j+u}^*\right|.
\end{eqnarray}
The following lemma from \cite{Schmidt2014} discloses the law of peak sidelobes in random binary sequences:
\begin{lemma}\cite{Schmidt2014}
For random binary sequences with length $N$, the following convergence holds
\begin{eqnarray}
&&P\left((\sqrt{2}-\epsilon)\sqrt{N\log N}\leq \mu_{\mathbf{x}} \leq (\sqrt{2}+\epsilon)\sqrt{N\log N}\right)\nonumber\\
&&\rightarrow 1,\qquad \mbox{as } N\rightarrow \infty,
\end{eqnarray}
for any $\epsilon>0$.
\end{lemma}
Therefore, when the sequence is normalized, the peak sidelobe level decreases as 
\begin{eqnarray}
\frac{\mu_{\mathbf{x}}}{\|\mathbf{x}\|^2}\approx\frac{\sqrt{2N\log N}}{N}  =\sqrt{\frac{2\log N}{N}},
\end{eqnarray}
which converges to zero as $N\rightarrow 0$. Therefore, a large bandwidth can drive the sidelobes to vanish.

\subsection{Imaging}
For imaging in (\ref{eq:imaging}), we turn it into the frequency domain by taking a Fourier transform and obtain
\begin{eqnarray}\label{eq:spectrum_imaing1}
Y(\mathbf{r},\omega)=A\int_{\Omega}J(\mathbf{r}')e^{-j\omega \tau(\mathbf{r},\mathbf{r}')}d\mathbf{r}'+N(\mathbf{r}),
\end{eqnarray}
where $\tau(\mathbf{r},\mathbf{r}')$ is the time delay between positions $\mathbf{r}$ and $\mathbf{r}'$. Here, we have implicitly assumed that the reflection coefficient is independent of frequency (thus being colorless) and the noise is white. Moreover, the phase shift term $e^{-j\omega \tau(\mathbf{r}')}$ can be incorporated into the reflection coefficient $J(\mathbf{r}')$, since imaging only involves the magnitude $|J(\mathbf{r}')|$ and omits the phase. Therefore, the received signal in (\ref{eq:spectrum_imaing1}) can be simplified to
\begin{eqnarray}\label{eq:spectrum_imaing2}
Y(\mathbf{r},\omega)=A\int_{\Omega}J(\mathbf{r}')d\mathbf{r}'+N(\mathbf{r}).
\end{eqnarray}
A direct attack on (\ref{eq:spectrum_imaing2}) is difficult, which involves the functional analysis of inverse problems, and the impact of bandwidth is unclear. For the simplicity of analysis, we assume the far field case, in which the Fraunhofer approximation can be employed to simplify the analysis. In particular, the source $J$ and the field measurement $Y$ (when there is no noise) form a pair of Fourier transform:
\begin{eqnarray}
Y(\mathbf{k})=\mathcal{F}(J(\mathbf{r}))+N(\mathbf{k}),
\end{eqnarray}
where $\mathbf{k}$ is the wave vector and the Fourier transform $\mathcal{F}$ sends the space-domain signal to the space of wave vectors.

\begin{figure}[!t]
\centering
\includegraphics[width=0.32\textwidth]{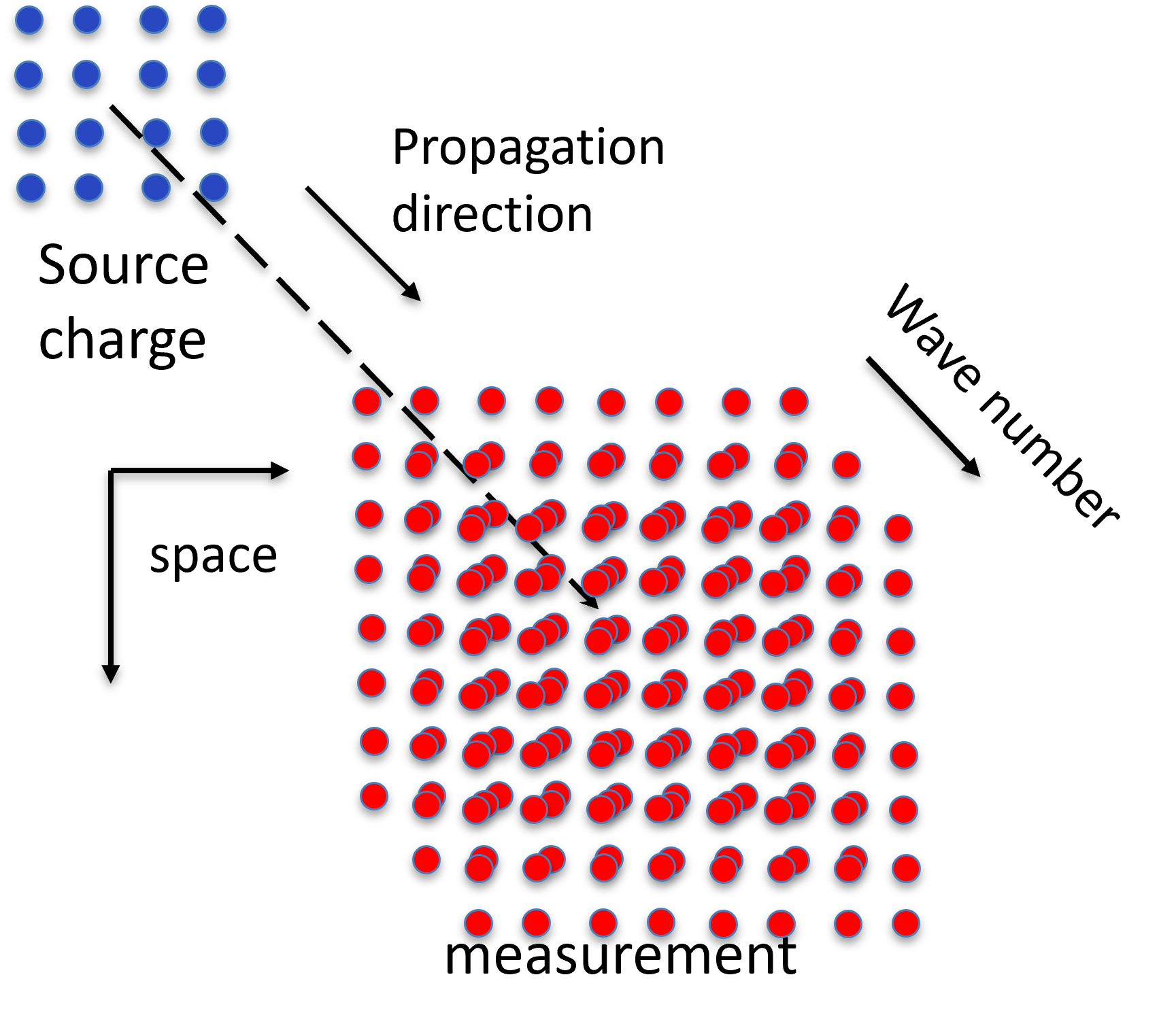}
\caption{An illustration of source charge and measurement}
\label{fig:propagationS}
\end{figure}

We can consider $J$ as the source charge. We further assume that the spatial bandwidth of the source charge is bounded, and a sampling over the grid $\{(m\delta L,n\delta L)\}_{m,n=-K,...,K}$ is sufficient to recover the source, where $\delta L$ is the sampling spacing. Therefore, the source charge $J$ can be represented by a $(2K+1)\times (2K+1)$ matrix $\mathbf{J}$. We assume that a planar aperture is used to measure the field, whose center is of distance $d$ to the center of the reflector. The measurement samples in the wavevector space are taken at $\{p\delta L, q\delta L\}_{p=1,...,N,q=1,...,N}$, as illustrated in Fig. \ref{fig:propagationS}. Therefore, the received signal is given by
\begin{eqnarray}
Y_k(p,q)=e^{-jkd}\sum_{m,n}e^{-jk\delta_k(pm+qn)\delta L^2}X(m,n)+N(m,n),
\end{eqnarray}
where $\delta_k=\frac{\delta_f}{c}$ is the spacing of the wave number, $k=\frac{f_c}{\delta_f},...,\frac{f_c+W}{\delta_f}$. One can understand $\frac{W}{\delta_f}$ as the number of subcarriers when OFDM is used for signaling.

We can incorporate the common phase factor $e^{-jkd}$ into the measurement $Y_k$. Then, the measurement can be rewritten in the following matrix form:
\begin{eqnarray}\label{eq:imaging_matrix}
\mathbf{Y}_k=\mathbf{F}_k\mathbf{X}\mathbf{F}_k^T+\mathbf{N},
\end{eqnarray}
where $k=K_1,...,K_2$ ($K_1=\frac{f_c}{\delta_f}$, $K_2=\frac{W}{\delta_f}$), and $(\mathbf{F}_k)_{pm}=e^{-jk\delta_kpm\delta L^2}$. Then, (\ref{eq:imaging_matrix}) can be further simplified into the following vector form:
\begin{eqnarray}\label{eq:imaging_vector}
vec(\mathbf{Y}_k)=\mathbf{F}_k\otimes \mathbf{F}_k^Tvec(\mathbf{X})+vec(\mathbf{N}),
\end{eqnarray}
where $\otimes$ is the Kronecker product and $vec$ is the operation of stacking the column vectors in the matrix into a larger vector.
In practice, the received signal in (\ref{eq:imaging_vector}) is contaminated by thermal noise $N$. 

It is easy to show that the following mean vector is a sufficient statistic for the estimation:
\begin{small}
\begin{eqnarray}
\bar{\mathbf{Y}}&=&\frac{1}{K_2-K_1}\sum_{k=K_1}^{K_2}vec(\mathbf{Y}_k)\nonumber\\
&=&\frac{1}{K_2-K_1}\sum_{k=K_1}^{K_2}\mathbf{F}_k\otimes \mathbf{F}_k^Tvec(\mathbf{X})+\text{noise}.
\end{eqnarray}
\end{small}
We notice that the signal-to-noise ratio (SNR) of each element in the estimation is then given by
$\frac{K}{\sigma_n^2}trace(\mathbf{H}\mathbf{H}^H)\times\frac{1}{K}\| vec(\mathbf{X})\|^2,$
where $\mathbf{H}=\frac{1}{K_2-K_1}\sum_{k=K_1}^{K_2}\mathbf{F}_k\otimes \mathbf{F}_k^T$. It converges to a bounded value, as $K_2-K_1\rightarrow\infty$. Therefore, different from the ranging task in sensing, but similarly to the channel capacity in communications, the imaging problem has a convergent but imperfect performance, as the bandwidth tends to infinity while the transmit power is bounded.

\section{Wideband ISAC Signaling}\label{sec:wideband}
In this section, we will study the signaling of ISAC in the wideband regime. We propose to use QPSK as the modulation, which is asymptotically optimal in the low SNR-regime (due to the limited power over a wide band). The PSD is then optimized for controlling the sidelobes. Then, we will study the duality of OTFS for justifying its applications in wideband ISAC.

\subsection{QPSK Modulation and Sidelobe Control}
As we have discussed above, given a sufficiently large bandwidth, the improvement in the communication channel capacity is marginal when modulation is also carried over the signal amplitude. Therefore, we propose the following simple scheme for ISAC: using the signal structure of OFDM, QPSK is used for the modulation on each subcarrier, while the power of each subcarrier is optimized for the performance of sensing. 

For optimizing the performance of sensing, we adjust the PSD in (\ref{eq:opt_MSE}) for minimizing the Cramer-Rao bound, subject an upper bound for the ISL. Note that the autocorrelation function of time-domain signal $x(t)$ and its PSD, both being continuous, form a Fourier transform pair. Therefore, a flatter PSD results in less ISL; in the ideal case when the PSD is a constant, the autocorrelation function becomes a Dirac delta function with no sidelobe. In our context of OFDM signaling structure, we can only manipulate the samples of PSD over finitely many subcarriers; and therefore, even if we set the same PSD for all subcarriers, the time-domain ISL is nonzero. In this study, we consider an OFDM symbol comprising $N_s$ subcarriers, each transmitting an average of $N_b$ bits. The time-domain signal samples are denoted by $\{ x_k \}_{k=1, \ldots, N_s}$, with a total transmit power of $P_t$. For sensing purposes, we define the aperiodic time-domain autocorrelation function as
\begin{eqnarray}
r(l)=\sum_{k=1}^{N_s-l}x_k^*x_{k+l},\qquad l=0,...,N_s-1.
\end{eqnarray}
which quantifies the signal's self-similarity at different lags. Sensing performance is assessed via the normalized ISL, given by
\begin{eqnarray}\label{eq:ISL}
ISL=\frac{\sum_{k=1}^{N_s-1}|r(l)|^2}{|r(0)|^2},
\end{eqnarray}
where $|r(0)|^2$ is the mainlobe power, and the goal is to minimize sidelobe amplitudes $|r(l)|$ for $l > 0$ to enhance sensing accuracy by reducing interference with weak targets.

For infinite-length signals, the autocorrelation function and PSD are Fourier transform pairs. A flat PSD, with power evenly distributed across frequencies, yields an autocorrelation approximating a Dirac delta function with minimal sidelobes. In our finite-length OFDM system with $N_s$ subcarriers, the spectrum is discrete, and the PSD is approximated by $|X[k]|^2$, the squared magnitude of the DFT of $x[k]$.

To justify flattening the PSD to reduce sidelobes, we analyze the circular autocorrelation of the finite sequence:
\begin{equation}
   r_c(l) = \sum_{k=1}^{N_s} x_k^* x_{(k + l) \mod N_s}, \quad l = 0, \ldots, N_s - 1. 
\end{equation}

According to the Wiener-Khinchin theorem~\cite{wiener1930generalized}, the DFT of $r_c(l)$ yields
\begin{equation}
   \text{DFT}\{ r_c(l) \} = |X[k]|^2, 
\end{equation}
showing that a uniform $|X[k]|^2$ makes $r_c(l)$ approximate a Kronecker delta, $r_c(l) \approx c \cdot \delta[l]$, minimizing sidelobes. In sensing, however, the aperiodic $r(l)$ is more relevant, corresponding to matched filter outputs. Therefore, we need to analyze the gap between the cyclic and acyclic autocorrelation functions.

\begin{figure}[!t]
\centering
\includegraphics[width=0.32\textwidth]{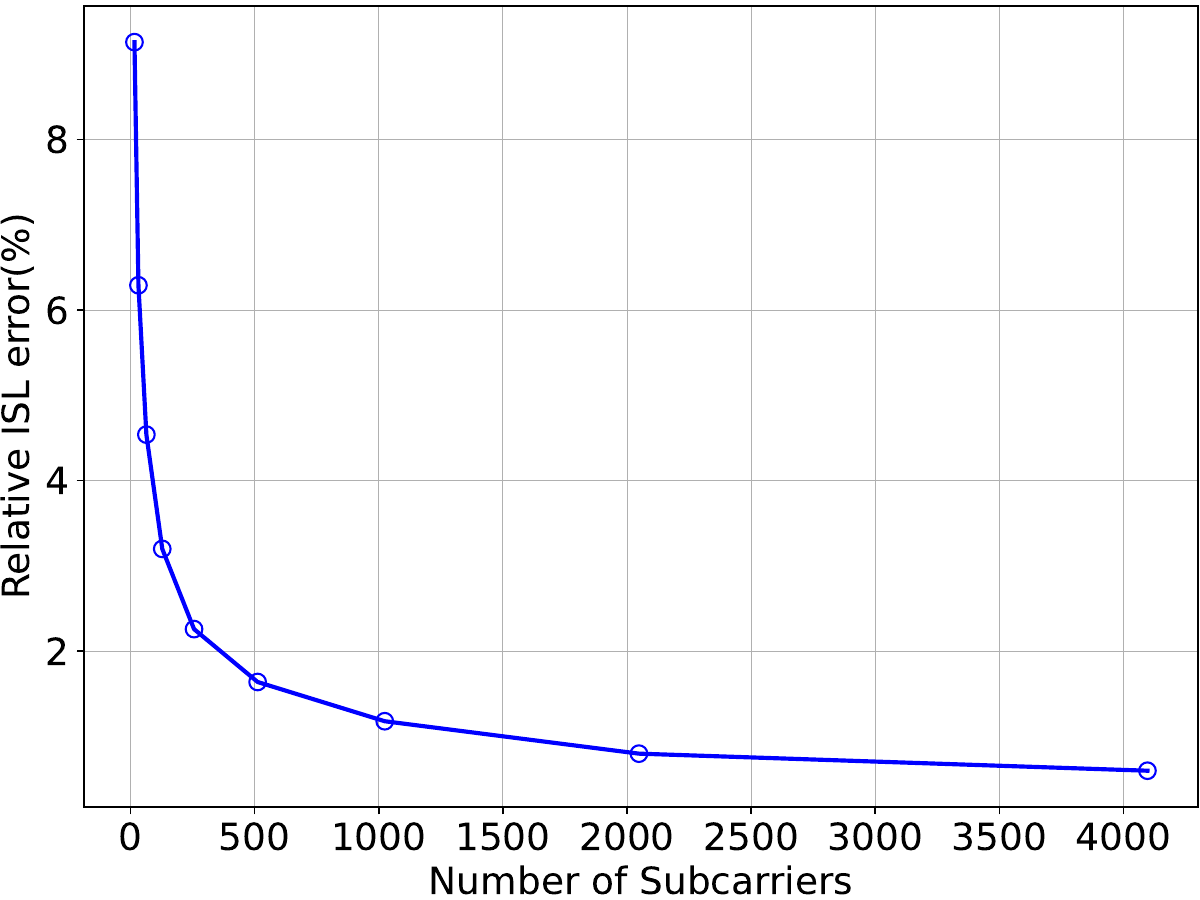}
\caption{ISL error between aperiodic correlation and periodic correlation versus different numbers of subcarriers. Above $N_s\approx 256$ the error falls below $2\%$, justifying the PSD‑variance surrogate used for large‑bandwidth systems.}
\label{fig:ISL_difference}
\end{figure}

\begin{figure*}[!t]
    \centering
    \subfigure[]
    {
        \label{fig:scheme_a}
        \includegraphics[width=1.55in]{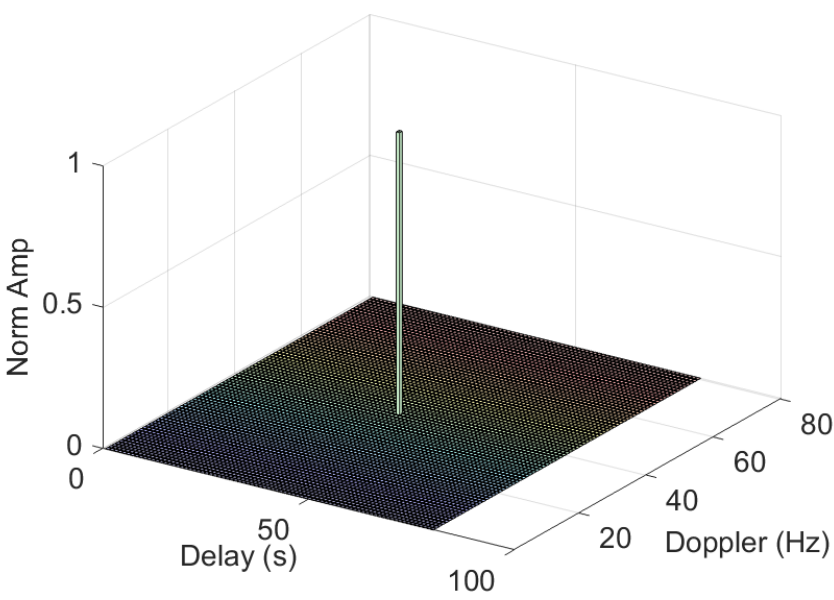}
    }
    \hspace{0.0001\linewidth}
    \subfigure[]
    {
       \label{fig:scheme_b}
        \includegraphics[width=1.55in]{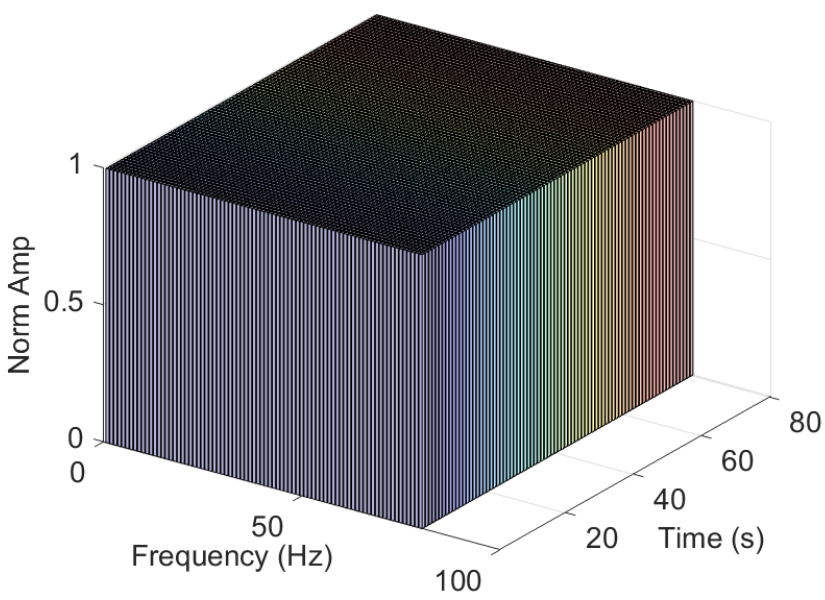}
    }
    \hspace{0.0001\linewidth}
    \subfigure[]
    {
        \label{fig:scheme_c}
        \includegraphics[width=1.55in]{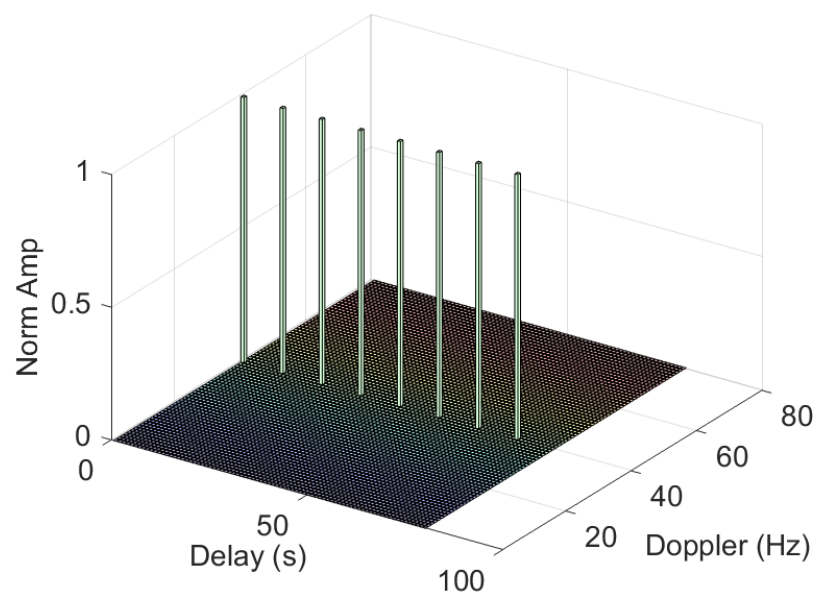}
    }
    \hspace{0.0001\linewidth}
    \subfigure[]
    {
        \label{fig:scheme_d}
        \includegraphics[width=1.55in]{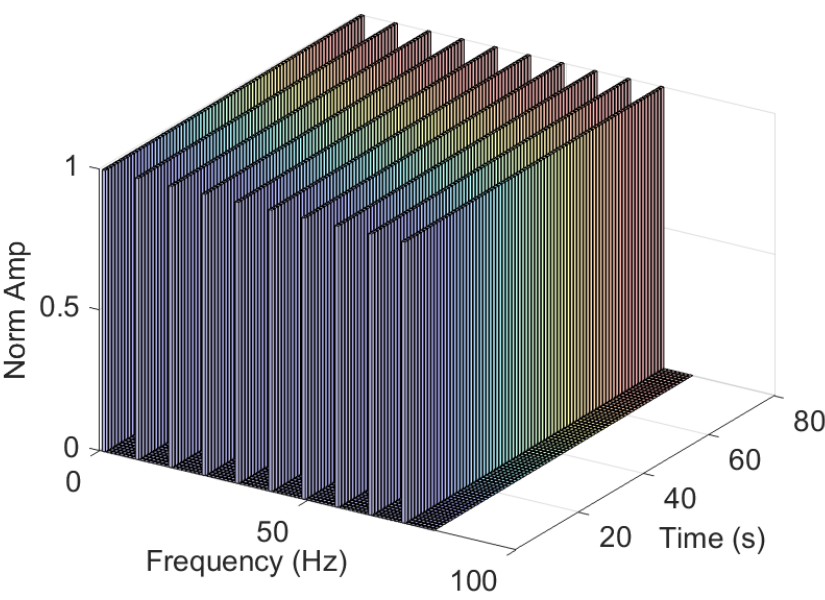}
    }
    
\caption{Representations of pilot signal in different domains: (a) signal in DD domain with single pilot, (b) corresponding signal to (a) in TF domain with single pilot, (c) signal in DD domain with multiple pilots along delay axis, (d) corresponding signal to (c) in TF domain with multiple pilots along delay axis.}
\label{fig:schemes}
\end{figure*}

To this end, we consider small lags $l \ll N_s$, especially with large $N_s$, $r(l) \approx r_c(l)$, as the periodic wrapping in $r_c(l)$ has minimal effect. The difference is given by
\begin{equation}
  r_c(l) - r(l) = \sum_{k=N_s - l + 1}^{N_s} x_k^* x_{k + l - N_s},  
\end{equation}
representing wrapped terms. The approximation $r(l) \approx r_c(l)$ introduces a bias, $r_c(l) - r(l)$, exact for infinite signals but imperfect for finite $N_s$. The bias magnitude is
\begin{equation}
   |r_c(l) - r(l)| = \left| \sum_{k=N_s - l + 1}^{N_s} x_k^* x_{k + l - N_s} \right|. 
\end{equation}
Using the Cauchy-Schwarz inequality, we have
\begin{equation}
   |r_c(l) - r(l)| \leq \sqrt{ \sum_{k=N_s - l + 1}^{N_s} |x_k|^2 \sum_{k=N_s - l + 1}^{N_s} |x_{k + l - N_s}|^2 }.
\end{equation}
Assuming uniform power allocation so that $\mathbb{E}[|x_k|^2]=P_t/N_s$ for the $N_s$ active subcarriers (with $P_t$ the total average transmit power), the sum over any $l$ terms is approximately $l\,P_t/N_s$. Therefore, we obtain
\begin{equation}
    |r_c(l) - r(l)| \leq \frac{l P_t}{N_s}.
\end{equation}

This bias is small relative to $r(0) = P_t$ for $l \ll N_s$, but grows with $l$, potentially reaching $P_t/2$ for $l \approx N_s/2$, affecting the ISL if sidelobes are not minimized.

For random data (e.g., QAM symbols), the expected bias $E[r_c(l) - r(l)] = 0$, as $E[x_k^* x_{k + l - N_s}] = 0$. However, the variance given by
\begin{equation}
  \text{Var}[r_c(l) - r(l)] = E\left[ \left| \sum_{k=N_s - l + 1}^{N_s} x_k^* x_{k + l - N_s} \right|^2 \right],  
\end{equation}
scales with $l \cdot (P_t / N_s)^2$, impacting individual ISL realizations. For large $N_s$, normalization by $|r(0)|^2 \approx P_t^2$ mitigates this, making a flat PSD a practical approach to minimize expected ISL. In Fig.~\ref{fig:ISL_difference}, we compare the calculated relative ISL error between the aperiodic correlation result (related to the PSD) and the periodic correlation result for various numbers of subcarriers. It can be observed that for large $N_s$, e.g., $N_s \ge 2024$, the ISL error is reduced to within 1\%. 

% \begin{figure}[!t]
% \centering
% \includegraphics[width=0.32\textwidth]{figures/corr_w_ISL_constraint.pdf}
% \caption{Periodic correlation results with different ISL constraints}
% \label{fig:CorrvsNs_w_ISL_constraints}
% \end{figure}

% In Fig. \ref{fig:CorrvsNs_w_ISL_constraints}, the correlation results under different $V_0$ constraints are visualized. It can be observed that as the $V_0$ constraint is relaxed from 0 to larger values, the sidelobe performance degrades. Asymptotically, when $V_0$ approaches infinity, the optimization is no longer constrained, leading to PSD convergence at spectral edges while degrading ISL performance. This, in turn, nullifies the radar's ranging functionality.

\subsection{Duality to OTFS}

As discussed in the previous section, the PSD largely determines the ISL: a flatter PSD produces a lower ISL, whereas a more uneven PSD increases it. Here, we extend this insight from OFDM to OTFS and show that it naturally carries over to delay estimation in the delay–Doppler (DD) domain.

Equations (\ref{eq:ISFFT}) and (\ref{eq:SFFT}) reveal that an OTFS transceiver is essentially an OFDM system augmented with an inverse SFFT at the transmitter input and a SFFT at the receiver output. Specifically, the delay-time (DT) domain representation $\mathbf{X}^{\mathrm{DT}}$ is derived by applying the inverse DFT to the DD domain signal $\mathbf{X}^{\mathrm{\mathrm{DD}}}$ along the Doppler axis. Similarly, the frequency-Doppler (FD) domain representation $\mathbf{X}^{\mathrm{FD}}$ is obtained by performing a DFT on $\mathbf{X}^{\mathrm{\mathrm{DD}}}$ along the delay axis; see \cite{OTFS_sub_Nyquist} for the full set of transformations.

Because these operations map the delay axis of the DD domain onto the frequency axis of the TF domain, enforcing a flat PSD in OTFS is equivalent to shaping the signal along the DD delay axis. Unlike OFDM, OTFS typically employs high-power embedded pilots in the DD domain for channel estimation \cite{embedded_pilot}. Fig.~\ref{fig:schemes} illustrates two pilot-placement strategies in DD domain and their TF counterparts. The single-pilot scheme in Fig. \ref{fig:scheme_a} maps, via the ISFFT, to a TF waveform whose subcarriers carry uniform power (Fig. \ref{fig:scheme_b}), yielding the flattest PSD and consequently the lowest ISL. By contrast, the multi-pilot configuration of Fig. \ref{fig:scheme_c} occupies several discrete delays; after the ISFFT it excites only a subset of subcarriers (Fig. \ref{fig:scheme_d}), producing a highly non-uniform PSD and a correspondingly higher ISL.

% \begin{figure}[!t]
% \centering
% \includegraphics[width=0.32\textwidth]{figures/OTFS_cdf_along_delay.pdf}
% \caption{CDFs of ISL for OTFS with different schemes}
% \label{fig:OTFS_cdf_along_delay}
% \end{figure}

% \begin{figure}[!t]
% \centering
% \includegraphics[width=0.32\textwidth]{figures/OTFS_cdf_along_Doppler.pdf}
% \caption{CDFs of ISL for OTFS with different schemes}
% \label{fig:OTFS_cdf_along_Doppler}
% \end{figure}

This behavior is equally intuitive from a delay-estimation perspective. A path delay manifests as a shift along the DD delay axis, and estimation reduces to row-wise cross-correlation with the pilot pattern. A single-pilot row yields one sharp correlation peak with minimal sidelobes; multiple pilots, however, introduce additional cross-terms that elevate sidelobes and degrade the ISL.

% To demonstrate this, we investigate an OTFS system characterized by the following parameters: delay bins ($M_{\tau} = 80$), Doppler bins ($N_{\nu} = 80$), and subcarrier spacing ($2.5$ MHz). The system employs rectangular pulse shaping. Here, $E_s = \mathbb{E}{ |\mathbf{X}_d^{\mathrm{DD}}[k,l]|^2 }$ denotes the average data symbol energy, and $E_p = \mathbb{E}{ |\mathbf{X}_p^{\mathrm{DD}}[k,l]|^2 }$ denotes the average pilot symbol energy. To facilitate a fair comparison, we fix the data and pilot powers at constants $E_s = 1$ and $E_p = 0.15$, respectively, while varying only the number of pilot symbols. When pilots are placed along the delay axis, the ISL results are shown in Fig.~\ref{fig:OTFS_cdf_along_delay}. We observe that increasing the number of pilots from 1 to 40 leads to significant degradation in ISL performance. This aligns with our earlier discussion that multiple pilot placements along the delay axis induce PSD variations in the TF domain, thereby degrading the ISL. Conversely, positioning pilots along the Doppler axis maintains a constant PSD across the frequency domain. The results depicted in Fig.~\ref{fig:OTFS_cdf_along_Doppler} indicate that ISL remains nearly unchanged in this scenario, validating our conclusion.

\section{Joint Optimization for Communication and Sensing}
\label{sec:joint_opt}

In Section~\ref{sec:wide_sensing}, we optimized the PSD over \( N \) subcarriers to enhance sensing performance, specifically targeting the ranging MSE and the ISL. The optimization maximized \( \sum_{n=1}^N X_n \omega_n^2 \) to improve ranging accuracy while constraining the variance of \( X_n \) to control ISL, under a total power constraint. However, in ISAC systems, the waveform must simultaneously support communication, where the channel often exhibits frequency-selective fading due to real-world non-idealities. This fading introduces variations in the channel frequency response across subcarriers, impacting the SNR and, consequently, the communication performance. Consequently, a joint optimization of \( X_n \) is necessary to balance sensing objectives (ISL and ranging MSE) with communication performance (e.g., data rate). In this section, we propose a promising approach to address this challenge, extending the framework to incorporate communication under frequency-selective fading.

\subsection{Problem Formulation}

For optimizing the performance of sensing, we adjust the PSD in (\ref{eq:opt_MSE}) for minimizing the CRB, which is equivalent in maximzing $\sum_{n=1}^N \omega_n^2X_n$. In a Ricean channel model, each subcarrier~$n$
experiences a deterministic channel gain $|h_n|^2$, which
depends primarily on the path loss and any residual phase shift
across frequencies. For the communication metric, we let
\begin{align}
  \label{eq:rate}
  C\bigl(\{X_n\}\bigr)
    \;=\;
    \sum_{n=1}^N \log_{2}\Bigl( 1 + \tfrac{|h_n|^2 X_n}{N_0(\frac{B}{N})} \Bigr),
\end{align}
represent the achievable sum-capacity across all subcarriers,
assuming a narrow subcarrier bandwidth (such that each
subcarrier is approximately flat) and $N_0$ is the noise power
spectral density. $B$ denotes the total physical bandwidth, and therefore $\frac{B}{N}$ is the bandwidth for each subcarrier.

In ISAC, a joint optimization must reconcile these objectives. Water-filling optimizes communication but may produce a highly variable PSD, increasing ISL and potentially reducing \( \sum_{n=1}^N X_n \omega_n^2 \) if power is not allocated to higher frequencies beneficial for ranging. Conversely, the sensing-optimal \( X_n \) may not align with the channel conditions, degrading the communication performance. We propose a constrained optimization that prioritizes communication rate while ensuring the sensing performance meets the minimum requirements:
\begin{eqnarray}\label{eq:opt_problem}
\begin{aligned}
\text{max}_{X_n} \quad & \alpha \zeta \left( C(\{X_n\}) \right) 
+ (1 - \alpha) \left( \sum_{n=1}^{N} \omega_n^2 X_n \right) \\
\text{s.t.} \quad & \sum_{n=1}^{N} (X_n - \bar{P})^2 \leq V_0, \\
& \sum_{n=1}^{N} X_n = NP, \\
& X_n \geq 0 \quad \forall n,
\end{aligned}
\end{eqnarray}
where $\alpha \in [0,1]$ is a weighting factor balancing the communication and sensing objectives.  $V_0$ is a predetermined threshold for the ISL. $\zeta = \frac{S_{\text{max}}}{C_{\text{max}}}$ is the scaling factor, where \( C_{\text{max}} \) denote the maximum capacity achieved by water-filling (when \( \alpha = 1 \)), and \( S_{\text{max}} = NP\max_{n} \omega_n^2 \) denote the maximum sensing metric when all power is allocated to the subcarrier with the highest \( \omega_n^2 \).

\subsection{Proposed Two-stage Algorithm}\label{sec:tw0_stage_algo}

Solving \eqref{eq:opt_problem} directly can be done via the Karush-Kuhn-Tucker (KKT) conditions, which requires simultaneously handling two coupled constraints with a nonlinear, multi-objective cost. To simplify it, we propose a two-stage approach that decouples the problem into two simpler subproblems:

\begin{algorithm}[t!]
\caption{Modified water-filling for waveform optimization}\label{alg:alg1}
\begin{algorithmic}[1]
\State \textbf{Input:} CSI $\{h_n\}$, sensing weights $\{\omega_n^2\}$, trade-off parameter $\alpha$
\State \textbf{Initialize:} $\lambda_{\text{low}} = 0$, $\lambda_{\text{high}} = \max_n \left( \frac{\alpha |h_n|^2}{(1-\alpha) \omega_n^2} \right)$
\While{$\sum X_n - NP > \epsilon$}
    \State Calculate the intermediate power allocation:
    \[
    X_n^* = \frac{\alpha\zeta}{(1-\alpha) \omega_n^2 + \lambda} - \frac{N_0}{|h_n|^2}
    \]
    \State Apply the projection to meet variance constraint:
    \[
    X_n^{\text{temp}} = P + \frac{V_0}{\sqrt{\sum (X_n^* - P)^2}} (X_n^* - P)
    \]
    \State Update $\lambda$ via the bisection search
\EndWhile
\State \textbf{Output:} Final power allocation $\{X_n^{\text{final}}\}$
\end{algorithmic}
\end{algorithm}

\subsubsection{Unconstrained Trade-off Optimization}
We temporarily drop the variance constraint in \eqref{eq:opt_problem} and the following unconstrained problem:
   \begin{equation}
   \begin{aligned}
     &  \max_{X_n \ge 0}
   \;
   \alpha\zeta \sum_{n=1}^N \log_2\!\Bigl(1 + \tfrac{|h_n|^2\,X_n}{N_0\frac{B}{N}}\Bigr)
   \;+\;
   (1-\alpha)\sum_{n=1}^N \omega_n^2\,X_n
   \quad \\
   &\text{s.t.}
   \quad
   \sum_{n=1}^N X_n = N\,P.
   \end{aligned}  
   \label{eq:unconstrained-problem}
   \end{equation}
   This problem admits a water-filling-like solution.  We denote the resulting solution by $X_n^*$ in the subsequent discussion.

   Here we sketch the derivation of the water-filling-like solution for the unconstrained problem in \eqref{eq:unconstrained-problem}. For brevity, we denote $\gamma_n \triangleq \tfrac{|h_n|^2 N}{N_0 B}$.  The Lagrangian is given by
\begin{equation}
\begin{aligned}
    \mathcal{L}
\;=\;
\frac{\alpha \zeta}{\ln2} \sum_{n=1}^N \ln\bigl(1 + \gamma_n\,X_n\bigr)
\;+\;
(1-\alpha)\sum_{n=1}^N \omega_n^2\,X_n \\
\;+\;
\lambda\Bigl(\sum_{n=1}^N X_n - N\,P\Bigr).
\end{aligned}
\label{eq:lagrangian-stage1}
\end{equation}
Taking partial derivatives w.r.t.\ $X_n$ and setting them to zero, we obtain
\begin{equation}
\frac{\alpha \zeta}{\ln2} \,\frac{\gamma_n}{1+\gamma_n X_n}
\;=\;
(\alpha-1)\,\omega_n^2
\;-\;\lambda.
\label{eq:stage1-derivative-zero}
\end{equation}
Rearrange to solve for $X_n^*$, we have
\begin{align}
X_n^*
&=\;
\left[
\frac{\alpha \zeta}{( \alpha-1)\,\omega_n^2\ln2 - \lambda\ln2}
\;-\;\frac{1}{\gamma_n}
\right]_{+},
\label{eq:Xn-solution-waterfill}
\end{align}
where $[z]_+ = \max(z,0)$ ensures nonnegativity.

In classical water-filling (i.e., no $\omega_n^2$ term), we would have
\(
X_n
=
\bigl[\tfrac{1}{\lambda} - \tfrac{1}{\gamma_n}\bigr]_{+}.
\)
Comparing with \eqref{eq:Xn-solution-waterfill} shows that the term $(1-\alpha)\,\omega_n^2$ effectively reduces each subcarrier's gain if $\omega_n$ is small, and boosts it if $\omega_n$ is large.  Equivalently, one can define a modified channel magnitude:
\begin{equation}
|h_n|^\prime
\;\;=\;\;
\frac{|h_n|}{\omega_n}
\;\sqrt{\frac{\alpha \zeta}{\,1-\alpha\,}}
\quad
\text{(up to some shift by $\lambda$).}
\end{equation}
Hence subcarriers at higher frequencies (larger $\omega_n$) see an ``improved'' effective gain, encouraging more power allocation to aid sensing performance (ranging resolution).  We thus recover a water-filling-like rule, but with each subcarrier's channel gain replaced by an $\omega_n$-weighted quantity.

\subsubsection{Variance-Constrained Projection}

   If the unconstrained solution $X_n^*$ already satisfies 
   $\sum (X_n^* - P)^2 \le V_0,$ then we are done.  Otherwise, project $X^*$ onto the feasible set defined by \eqref{eq:opt_problem}.  In practice, you can do an $\ell_2$-projection:
   \[
     X_n^\text{final}
     \;=\;
     P
     \;+\;
     \sqrt{\frac{V_0}{\sum_{k=1}^N (X_k^* - P)^2}}
     \;\bigl(X_n^* - P\bigr).
   \]
   This ensures $\sum (X_n^\text{final} - P)^2 = V_0$ and preserves $\sum X_n^\text{final} = \sum X_n^* = NP$.

The final water-level adjustment is performed via $\lambda$. The constant $\lambda$ is found by enforcing the total-power constraint.  Numerically, one typically uses a bisection or Newton method, adjusting $\lambda$ until $\sum X_n = N P$. The detailed algorithm is described in Algorithm~\ref{alg:alg1}.

Direct KKT-based methods require iterative computations to determine an additional dual variable (the Lagrange multipliers), which can introduce the numerical instability. By reformulating the problem into a two-stage subproblem, each stage becomes computationally straightforward. Stage 1 corresponds to a standard water-filling-like problem, while Stage 2 involves an Euclidean projection with a closed-form solution. Intuitively, Stage 1 determines the optimal allocation in the absence of sidelobe constraints, whereas Stage 2 adjusts the PSD to ensure compliance with the ISL constraint, $\sum (X_n-P)^2 \le V_0$. If the required projection is small, the solution remains close to the unconstrained optimum. In many practical cases, the Stage 1 solution either (a) already satisfies constraint \eqref{eq:opt_problem} or (b) requires only minor adjustments. Consequently, this approach achieves a near-optimal performance with significantly reduced computational complexity.

\subsection{Complexity Analysis}
\label{subsec:complexity}

In this subsection, we analyze the computational complexity of the proposed two-stage solution presented in Section~\ref{sec:tw0_stage_algo}. We compare it with: 1) the traditional single-stage method (e.g., interior-point or gradient-based) for the entire joint optimization, and 2) the exhaustive search. 

\begin{figure}[!t]
\centering
\includegraphics[width=0.31\textwidth]{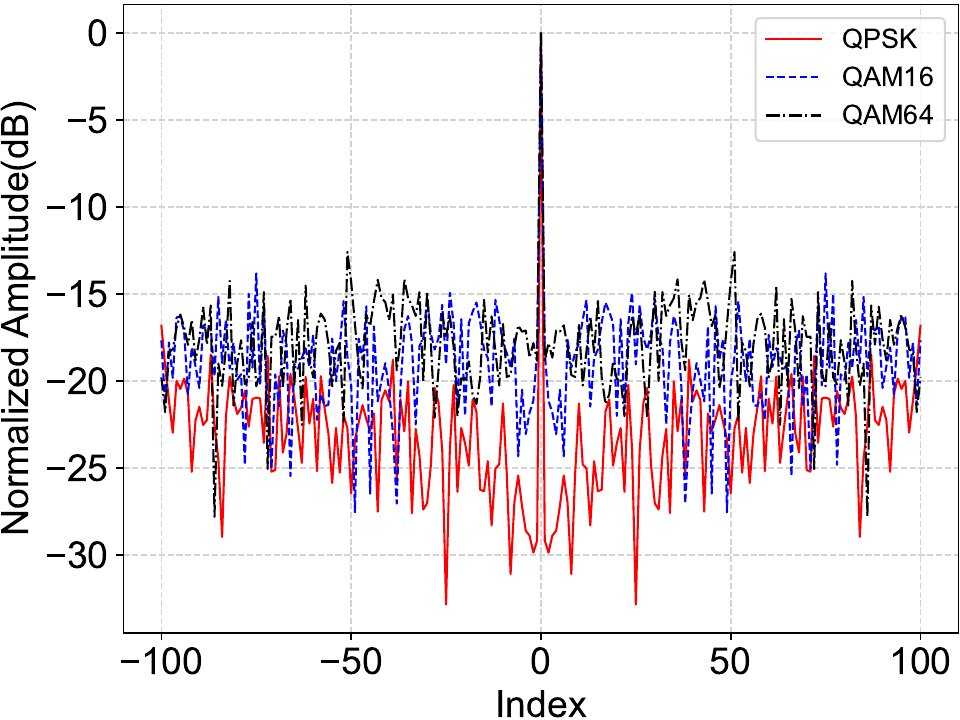}
\caption{Autocorrelation results of signals with different modulation orders.}
\label{fig:QPSK_QAM_corr_simulation}
\end{figure}

\begin{figure}[!t]
\centering
\includegraphics[width=0.35\textwidth]{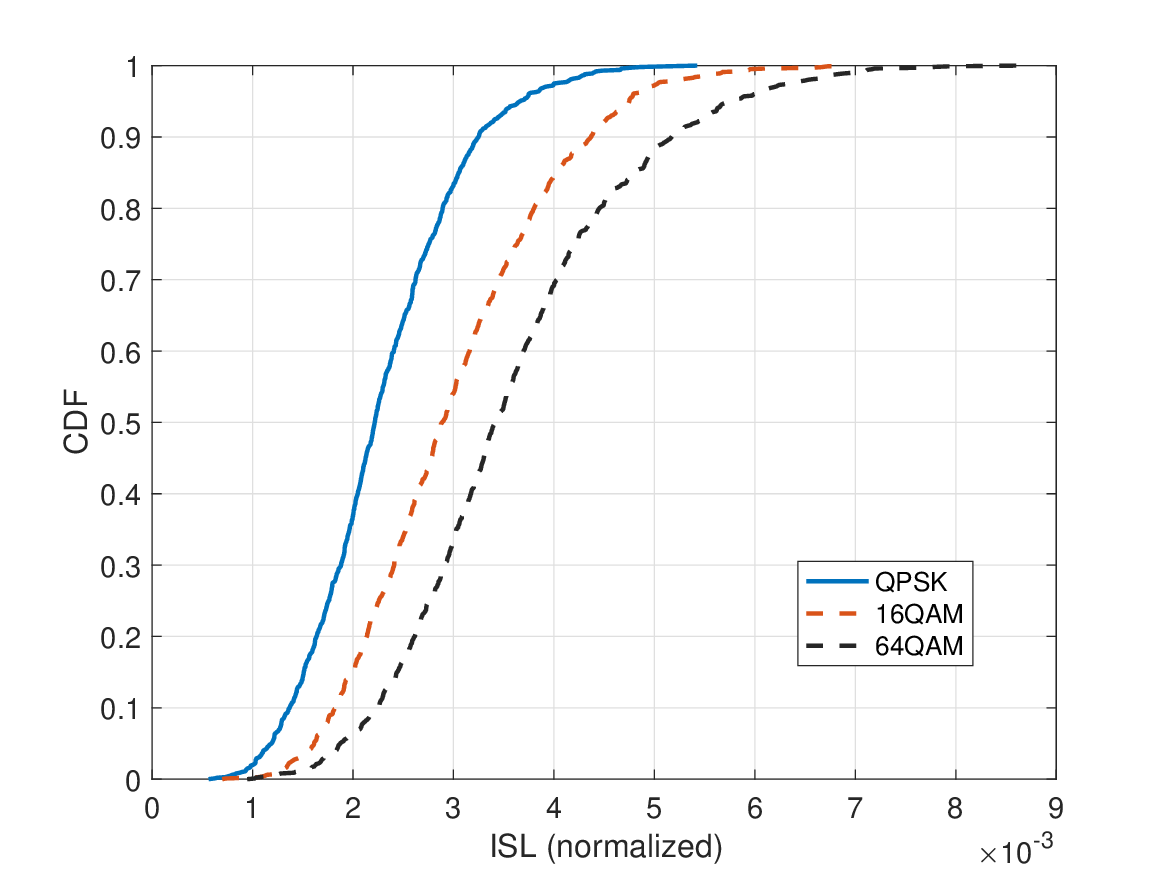}
\caption{CDFs of ISL for QSPK with constant PSD}
\label{fig:ISL}
\end{figure}

\subsubsection{Proposed Two-Stage Algorithm}

\paragraph{Stage~1 Complexity}
To solve \eqref{eq:unconstrained-problem}, one typically performs a one-dimensional bisection search on a Lagrange multiplier~$\lambda$.
Each bisection iteration has $\mathcal{O}(N)$ cost, and the search converges in $\mathcal{O}\bigl(\log(1/\epsilon)\bigr)$ iterations for precision~$\epsilon$. Hence, Stage~1 has a complexity $\mathcal{O}\Bigl(N \,\log\frac{1}{\epsilon}\Bigr)$.

\paragraph{Stage~2 Complexity}
If 
\(
\sum_{n=1}^N (X_n^\star - P)^2 \,\le\, V_0
\)
already holds, no projection is needed. Otherwise, we must solve 
\[
\begin{aligned}
&\min_{\{X_n\}} \sum_{n=1}^N \bigl(X_n - X_n^\star\bigr)^2, 
\\
&\text{s.t.}
\quad
\sum_{n=1}^N (X_n - P)^2 \;=\; V_0,
\quad
\sum_{n=1}^N X_n \;=\; N\,P.
\end{aligned}
\]
This is an $\ell_2$-projection onto the intersection of a sphere and a hyperplane. A closed-form or bisection-based approach finds $\{X_n\}$ in $\mathcal{O}(N)$. 

Hence, with computational cost of the total complexity for the proposed two-stage algorithm is approximately given by 
\begin{equation}
   \mathcal{O}\Bigl(N\,\log\frac{1}{\epsilon}\Bigr) \;+\; \mathcal{O}(N) 
\;\approx\;
\mathcal{O}\Bigl(N\,\log\frac{1}{\epsilon}\Bigr). 
\end{equation}

\begin{figure}[!t]
\centering
\includegraphics[width=0.35\textwidth]{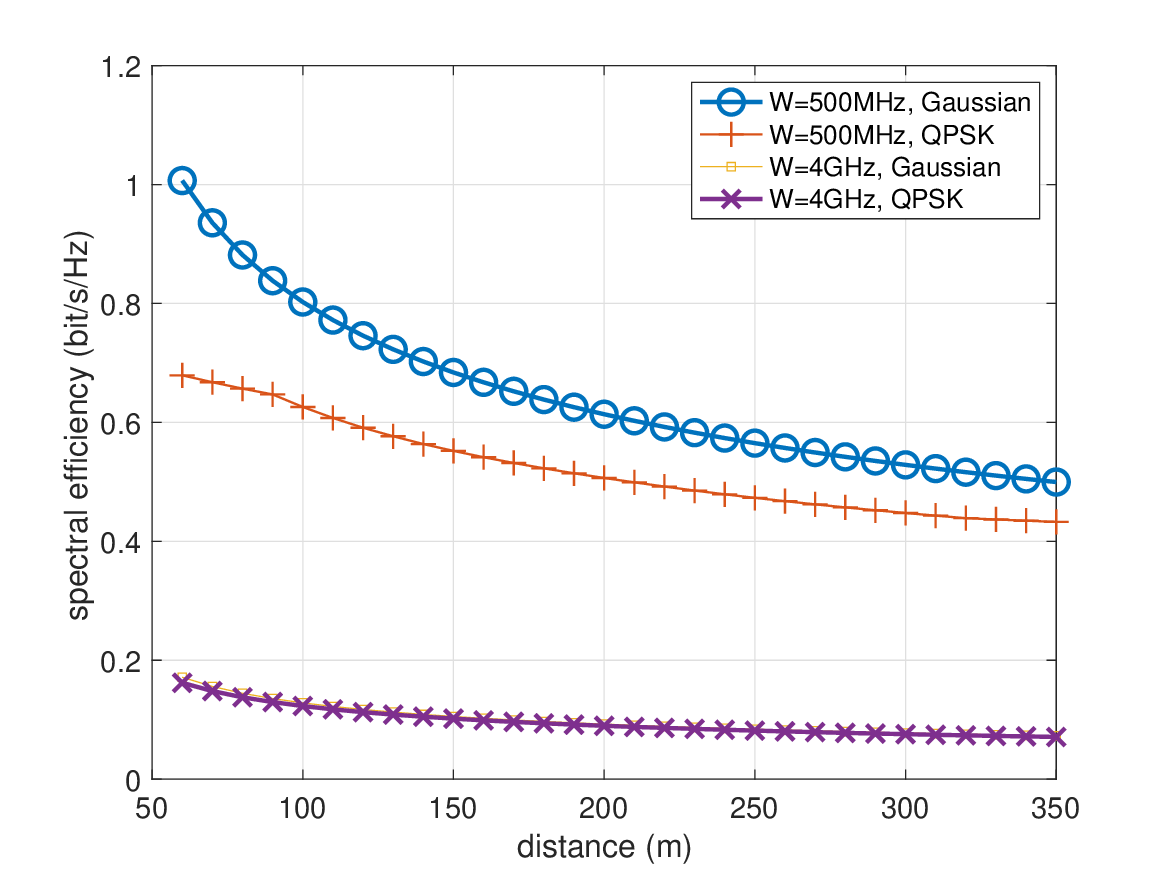}
\caption{Spectral efficiency of Gaussian signaling and QPSK}
\label{fig:SE}
\end{figure}

\begin{figure}[!t]
\centering
\includegraphics[width=0.35\textwidth]{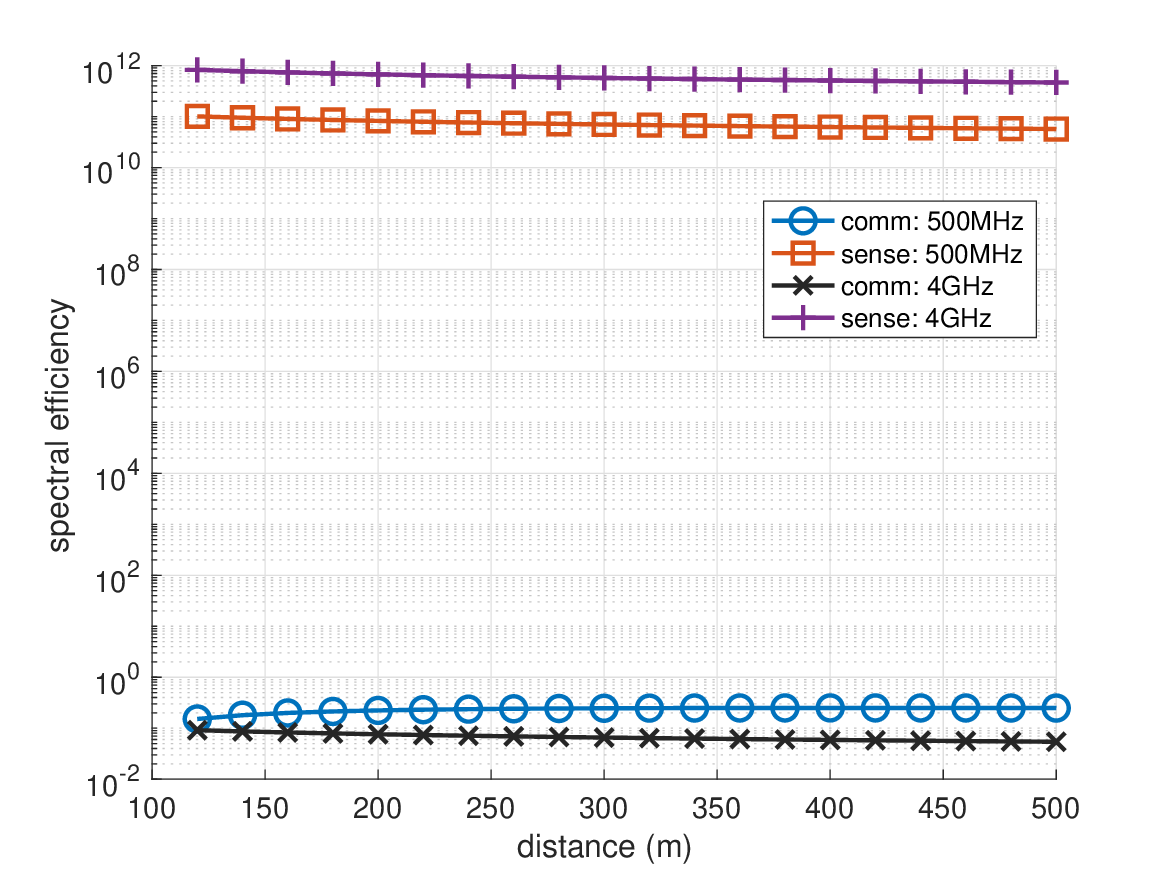}
\caption{Spectral efficiencies of communications and sensing}
\label{fig:SE_both}
\end{figure}

\subsubsection{Traditional Single-Stage Methods}

One could instead solve the original problem (including the sidelobe variance constraint) in a single shot via, e.g., interior-point or gradient-based methods. Such methods jointly handle the non-linear objective. Interior-point approaches often exhibit $\mathcal{O}(N^2)$ or worse complexity for a large \(N\). In wideband systems with thousands of subcarriers, this can become computationally expensive compared to the linear-scaling approach of our two-stage method.

\subsubsection{Exhaustive Search}

Performing a brute-force search over all possible power allocations on each of $N$ subcarriers (even when quantized to $L$ discrete values per subcarrier) incurs a complexity of $\mathcal{O}(L^N)$, which is intractable beyond very small $N$. Therefore, exhaustive search is typically not an option in large-scale systems.

% \begin{itemize}
% \item Proposed Two-Stage: 
%   Scales approximately linearly in the number of subcarriers ($\mathcal{O}(N \log(1/\epsilon))$). It first solves a "modified water-filling'' problem, then enforces sidelobe constraints via a low-cost projection. 
% \item Traditional Methods: 
%   Single-stage interior-point or gradient-based algorithms often have $\mathcal{O}(N^2)$ complexity or higher. 
% \item Exhaustive Search: 
%   Exhibits exponential $\bigl(\mathcal{O}(L^N)\bigr)$ blow-up, making it impractical. 
% \end{itemize}

These comparisons show that the two-stage solution strikes a favorable balance. This approach is significantly more scalable than single-stage or exhaustive methods, making it promising for wideband ISAC deployments with large $N$.

\begin{figure}[!t]
\centering
\includegraphics[width=0.35\textwidth]{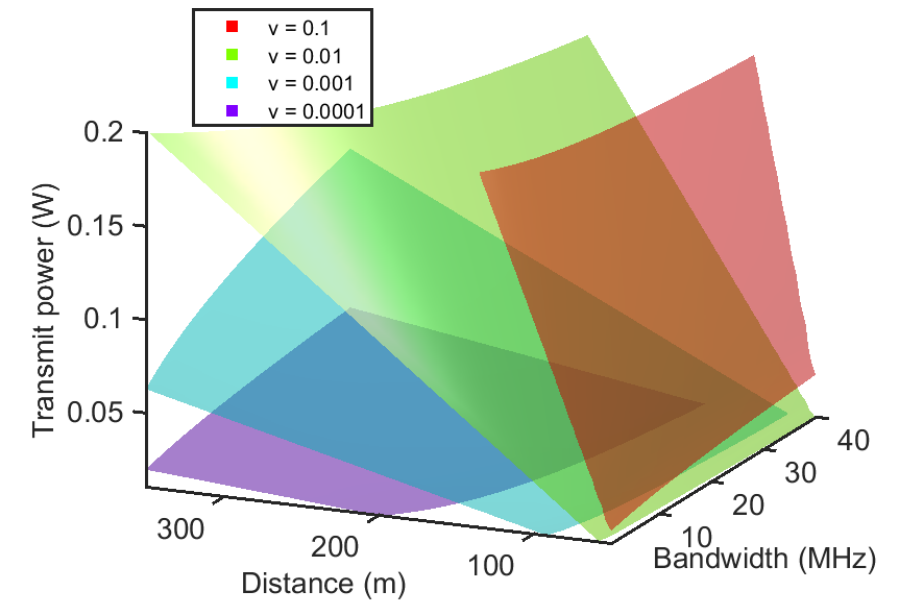}
\caption{Scope where QPSK and Gaussian has minor error gap}
\label{fig:surfaces}
\end{figure}

\section{Numerical and Experimental Results}\label{sec:numerical}
In this section, we provide numerical results to demonstrate the conclusions drawn in this paper. We focus on the proposed QPSK+OFDM ISAC scheme.

\subsection{ISL Distribution}

In Fig.~\ref{fig:QPSK_QAM_corr_simulation}, we visually present the periodic correlation results for QPSK, 16QAM, and 64QAM. It is clear that QPSK exhibits the lowest sidelobe levels, followed by 16QAM, whereas 64QAM shows the most pronounced sidelobes.

In Fig. \ref{fig:ISL}, we plotted the cumulative distribution functions (CDFs) of the ISL (normalized by $r^2(0)$) for QPSK, 16QAM and 64QAM. We observe that maintaining a constant PSD in QPSK substantially reduces the ISL. For a higher order of QAM, the ISL is increased due to the greater variance in the PSD and thus the higher ISL in the time domain. 

\subsection{Degradation of Communication Spectral Efficiency}
In Fig. \ref{fig:SE}, we plotted the spectral efficiencies of Gaussian signaling, which achieves the maximal channel capacity, and the OFDM+QPSK scheme. We assume that the total bandwidth $W$ is 500MHz and 4GHz, respectively. The transmit power is assumed to be 0.2W. The thermal noise PSD is set to -150dBm/Hz. The pathloss model is $L=48.6+3.5\log_{10}d$ (dB), where $d$ is the distance (in the unit of meter) between the transmitter and receiver. We let the distance $d$ range from 50m to 350m. The calculation of the QPSK channel capacity can be found in the proof of Theorem 11 in \cite{Verdu2002}. We observe that there is still a significant gap between the Gaussian signaling and QPSK scheme, when the bandwidth is 500MHz, although the gap substantially decreases as the distance $d$ becomes larger. When the bandwidth is 4GHz, the spectral efficiencies of the two schemes become indistinguishable, while they are both lower than the case of 500MHz bandwidth. The corresponding spectral efficiencies are plotted in Fig. \ref{fig:SE_both}.

\begin{figure}[!t]
\centering
\includegraphics[width=0.31\textwidth]{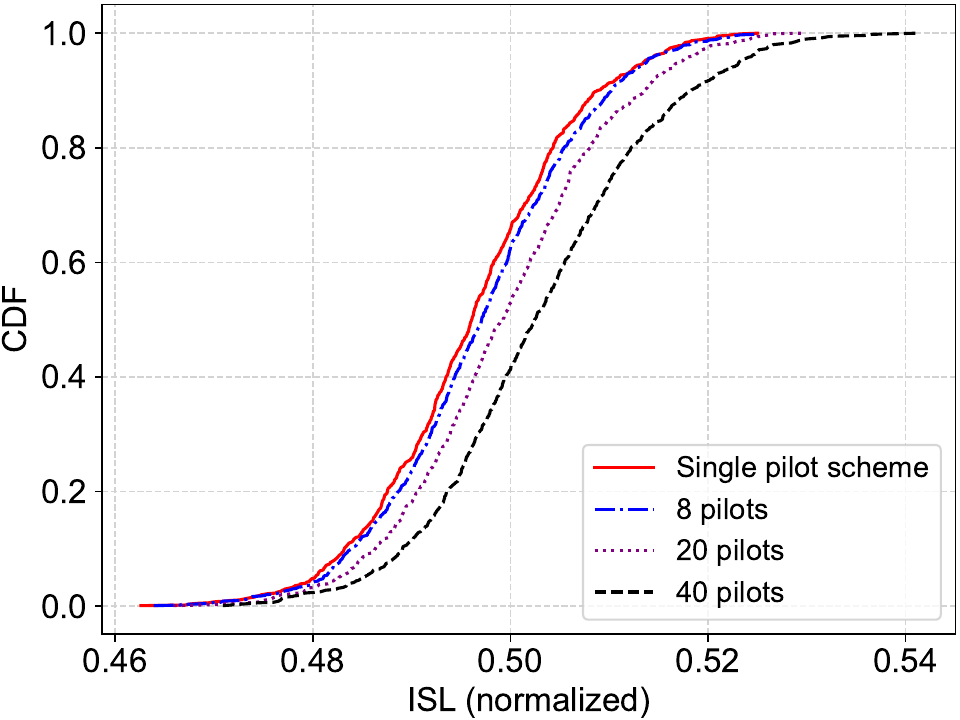}
\caption{CDFs of ISL for OTFS with different schemes}
\label{fig:OTFS_cdf_along_delay}
\end{figure}

\begin{figure}[!t]
\centering
\includegraphics[width=0.31\textwidth]{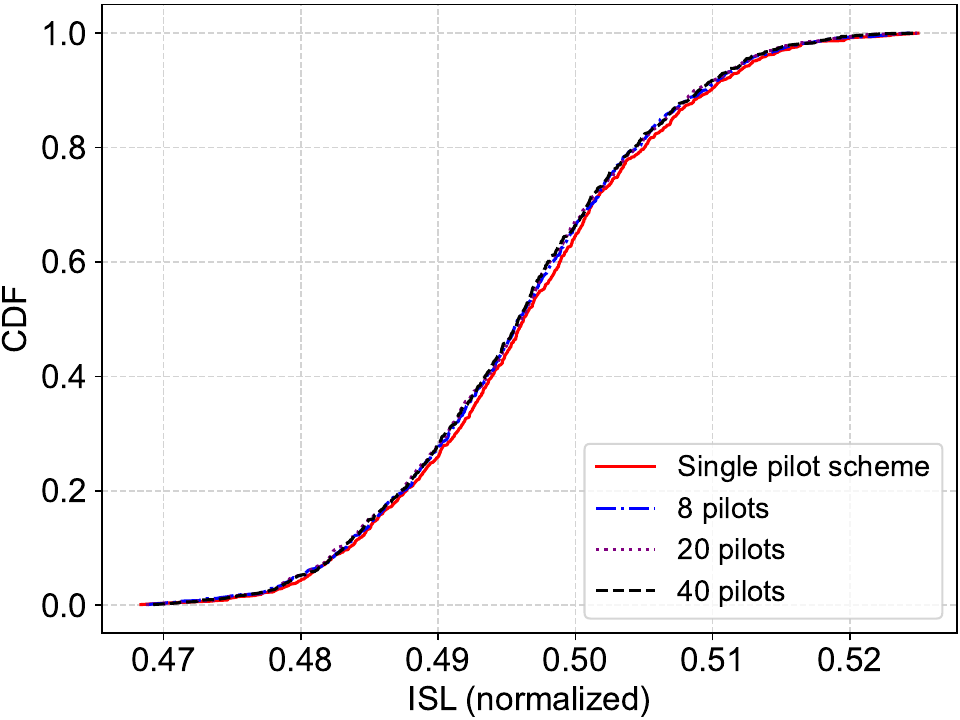}
\caption{CDFs of ISL for OTFS with different schemes}
\label{fig:OTFS_cdf_along_Doppler}
\end{figure}

Moreover, we identify the region where QPSK achieves communication efficiency comparable to Gaussian signaling, considering transmit power, distance, and bandwidth. The thermal noise PSD is fixed at $-150$~dBm/Hz. Transmit power varies from 0.01 to 0.2 W, distance ranges from 50 to 350 meters, and bandwidth spans from 100 MHz to 4 GHz. To quantify how close QPSK is to the Gaussian benchmark, we define a normalized gap $v$: the percentage by which QPSK’s efficiency falls short of Gaussian’s. The boundary surfaces representing regions where the gap between QPSK and Gaussian signaling is within a specified tolerance $v$ are shown in Fig.~\ref{fig:surfaces}. We observe that larger tolerance values ($v$) correspond to wider regions. Conversely, with a stringent tolerance of $v = 0.0001$, the region shrinks significantly, requiring transmit power below 0.05 W and distances generally exceeding 200 m. An increase in distance slightly relaxes the bandwidth requirement; for a bandwidth of 100 MHz, the required distance exceeds 200 m, whereas for bandwidths approaching 4 GHz, distances over 100 m are sufficient. This result visually illustrates the operational conditions under which QPSK approaches optimal Gaussian signaling performance.

\begin{figure*}[!h]
    \centering
    \subfigure[Power allocation with $\alpha=1$.]
    {
        \label{fig:power_alpha1}
        \includegraphics[width=1.9in]{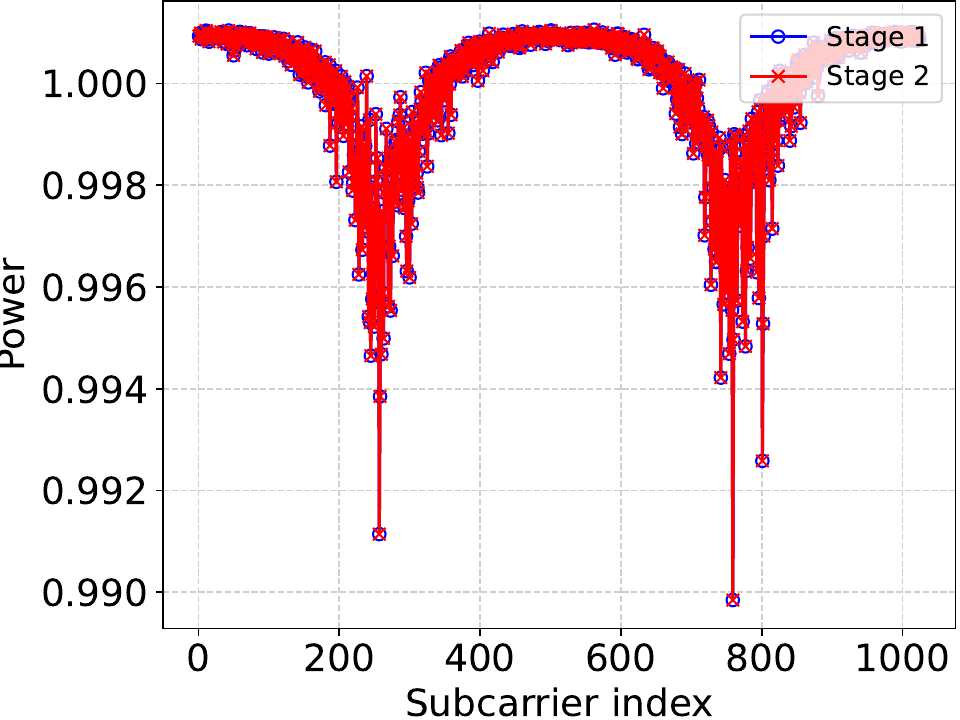}
    }
    \hspace{0.01\linewidth}
    \subfigure[Power allocation with $\alpha=0.5$.]
    {
        \label{fig:power_alpha5}
        \includegraphics[width=1.9in]{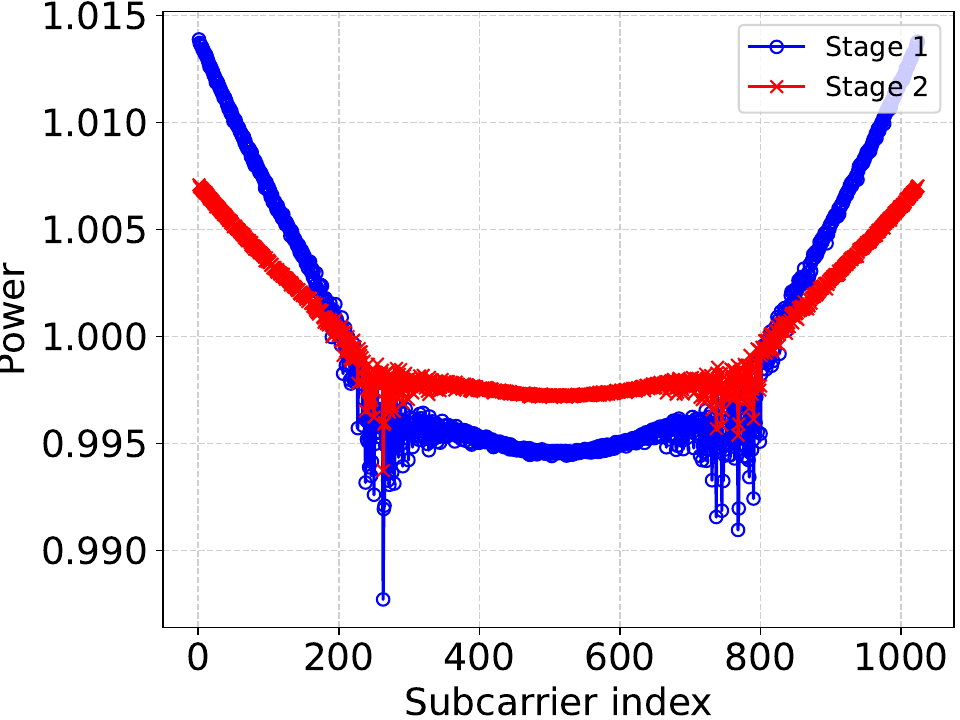}
    }
    \hspace{0.01\linewidth}
\subfigure[Power allocation with $\alpha=0$.]
    {
        \label{fig:power_alpha0}
        \includegraphics[width=1.9in]{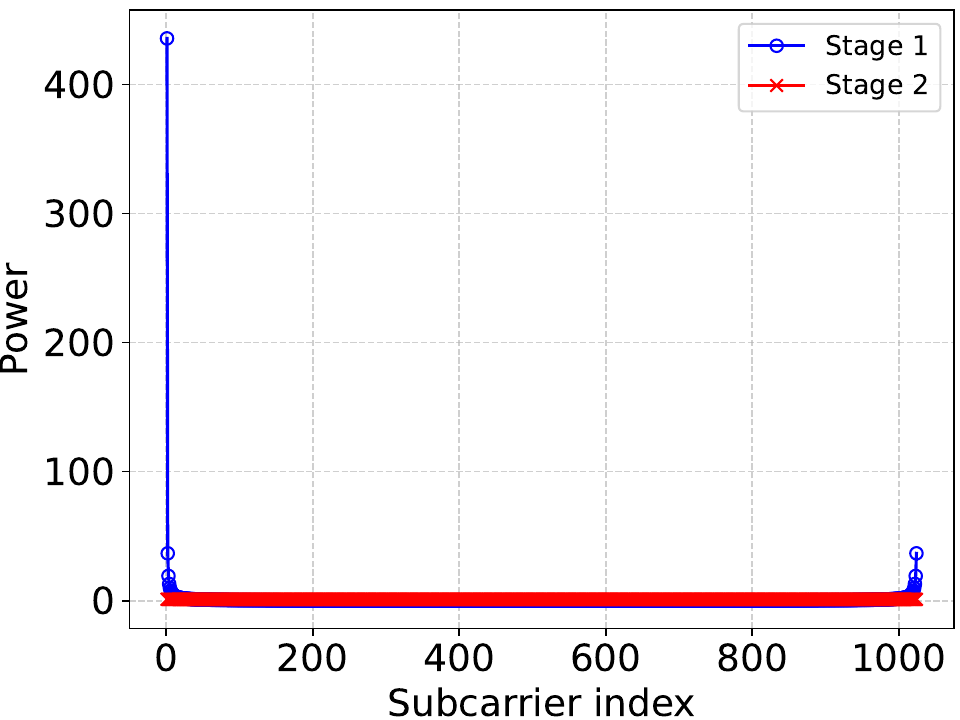}
    }
    \hspace{0.01\linewidth}
\caption{Power allocation with different weighting factor $\alpha$.}
\label{fig:power_alpha}
\end{figure*}

\begin{figure*}[!h]
    \centering
    \subfigure[Correlation result with $\alpha=1$.]
    {
        \label{fig:corr_alpha1}
        \includegraphics[width=1.9in]{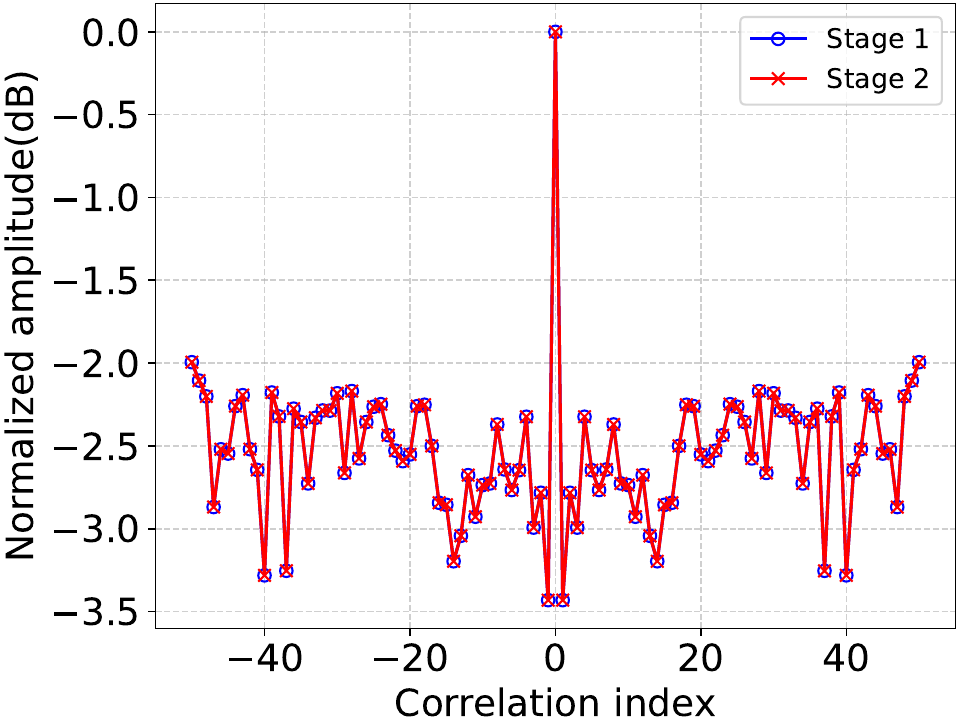}
    }
    \hspace{0.01\linewidth}
    \subfigure[Correlation result with $\alpha=0.5$.]
    {
        \label{fig:corr_alpha5}
        \includegraphics[width=1.9in]{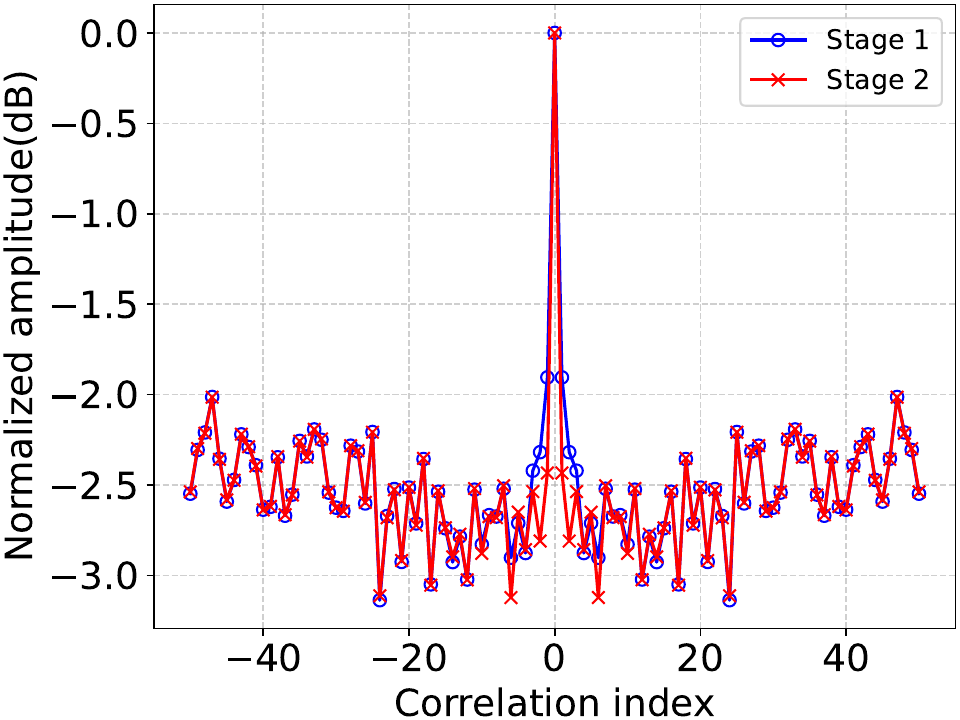}
    }
    \hspace{0.01\linewidth}
\subfigure[Correlation result $\alpha=0$.]
    {
        \label{fig:corr_alpha0}
        \includegraphics[width=1.9in]{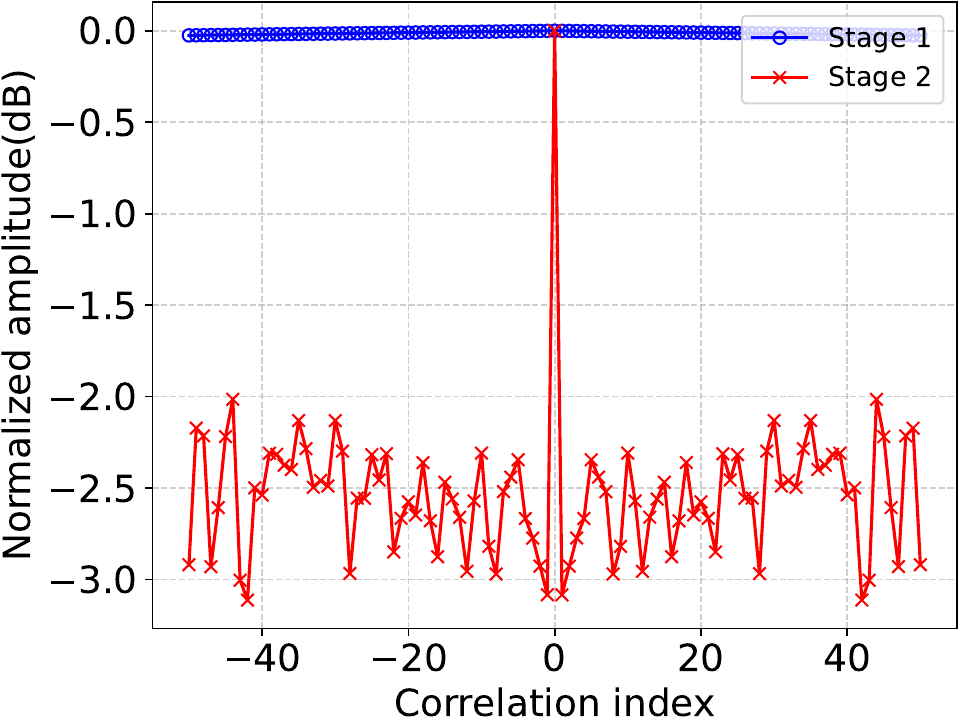}
    }
    \hspace{0.01\linewidth}
\caption{Correlation results with different weighting factor $\alpha$.}
\label{fig:corr_alpha}
\end{figure*}

\subsection{OTFS ISL Control}

To demonstrate the ISL reduction in OTFS via PSD control, we investigate an OTFS system characterized by the following parameters: delay bins ($M_{\tau} = 80$), Doppler bins ($N_{\nu} = 80$), and subcarrier spacing ($2.5$ MHz). The system employs rectangular pulse shaping. Here, $E_s = \mathbb{E}{ |\mathbf{X}_d^{\mathrm{DD}}[k,l]|^2 }$ denotes the average data symbol energy, and $E_p = \mathbb{E}{ |\mathbf{X}_p^{\mathrm{DD}}[k,l]|^2 }$ denotes the average pilot symbol energy, where $\mathbf{X}_d^{\mathrm{DD}}$ and $\mathbf{X}_p^{\mathrm{DD}}$ denote the data symbols and pilot symbols in DD domain. To facilitate a fair comparison, we fix the data and pilot powers at constants $E_s = 1$ and $E_p = 0.15$, respectively, while varying only the number of pilot symbols. When pilots are placed along the delay axis, the ISL results are shown in Fig.~\ref{fig:OTFS_cdf_along_delay}. We observe that increasing the number of pilots from 1 to 40 leads to significant degradation in ISL performance. This aligns with our earlier discussion that multiple pilot placements along the delay axis induce PSD variations in the TF domain, thereby degrading the ISL. Conversely, positioning pilots along the Doppler axis maintains a constant PSD across the frequency domain. The results depicted in Fig.~\ref{fig:OTFS_cdf_along_Doppler} indicate that ISL remains nearly unchanged in this scenario, validating our conclusion.

\subsection{Joint Sensing and Communication Optimization}

In this subsection, we present numerical results to validate the effectiveness of our proposed two-stage optimization approach. We assume that the signal bandwidth is 1GHz. The number of OFDM subcarriers is 1,024, The transmit power is assumed to be 0.2W. Channel attenuation is 50~dB. The thermal noise PSD is set to $-150$~dBm/Hz. The tolerable ISL variance is set as $V_0 = P^2$. To demonstrate adaptability to frequency-selective channels, a synthetic slow channel fading profile with two attenuation notches at subcarriers 260 and 760 is designed, intentionally creating non-uniform channel conditions for visualizing power allocation dynamics. Small-scale (fast) fading is modeled as a Rayleigh fading process, with the complex fading coefficients following a zero-mean circularly symmetric complex Gaussian distribution.

\subsubsection{Two-stage Optimization Visualization}

 To illustrate the power allocation strategy of our proposed two-stage algorithm, we present the results in Fig.~\ref{fig:power_alpha} for different weighting factors: $\alpha = 1, 0.5, 0$. The parameter $\alpha$ determines the balance between communication and sensing performance:

\begin{itemize}
    \item When $\alpha = 1$, the optimization prioritizes the communication performance.
    \item When $\alpha=0$, the optimization is fully dedicated to enhancing the sensing performance.
    \item When $\alpha = 0.5$, both communication and sensing objectives are considered.
\end{itemize}

Fig.~\ref{fig:power_alpha1} shows the power allocation results when $\alpha = 1$. In this case, the algorithm behaves similarly to traditional water-filling, ensuring that the ISL constraint is not violated. Consequently, the outputs of Stage 1 and Stage 2 remain the same, meaning that no additional projection is required.

When $\alpha = 0.5$, as illustrated in Fig.~\ref{fig:power_alpha5}, the algorithm starts balancing both sensing and communication objectives. Stage 1 solution allocates more power to the spectral edges to improve the sensing resolution while simultaneously reducing power in poor subcarriers to maintain communication performance. However, this allocation leads to ISL constraint violations because excessive power concentration at the spectral edges results in a non-flat PSD. To rectify this, the Stage 2 projection step is activated to redistribute power and enforce the ISL constraint.

Fig.~\ref{fig:power_alpha0} presents the case when $\alpha = 0$, where the optimization is purely focused on sensing. Here, Stage 1 solution heavily concentrates power at the spectral edges, leaving most other subcarriers with minimal power. This aggressive allocation significantly violates the ISL constraint, necessitating Stage 2 projection to flatten the PSD and restore compliance.

To further analyze the impact of these adjustments, Fig.~\ref{fig:corr_alpha} compares the correlation results before and after Stage 2 projection. As seen in Fig.~\ref{fig:corr_alpha1}, when $\alpha = 1$, the correlation remains unchanged since the ISL constraint is naturally satisfied. However, in Figures~\ref{fig:corr_alpha5} and \ref{fig:corr_alpha0}, where the ISL constraint is violated in Stage 1, Stage 2 projection strictly adjusts the power allocation to ensure compliance, thereby modifying the correlation patterns accordingly.

\begin{figure}[!t]
\centering
\includegraphics[width=0.32\textwidth]{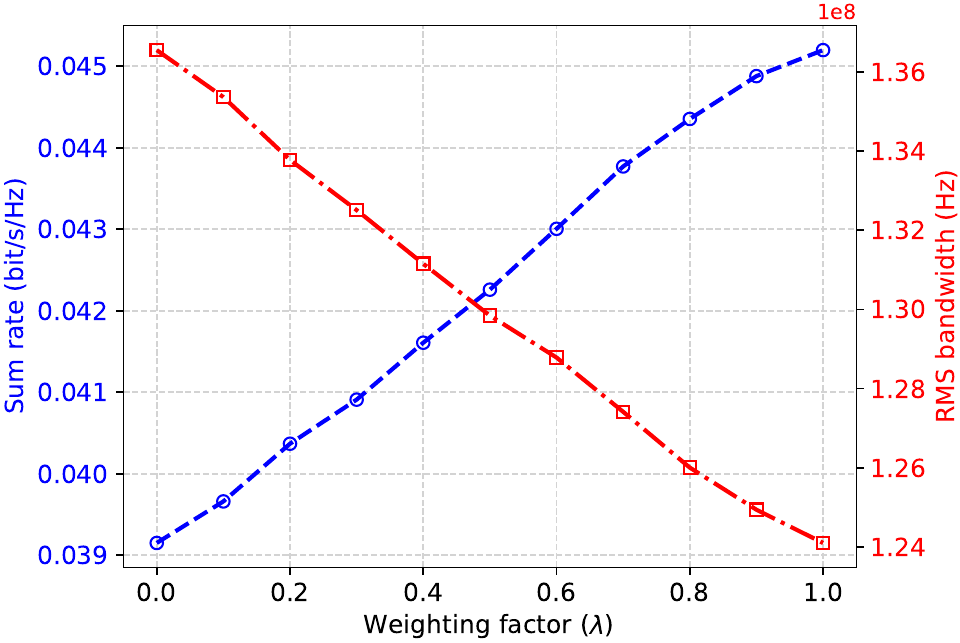}
\caption{Spectral efficiencies of communications and sensing}
\label{fig:tradeoff}
\end{figure}

\begin{figure}[!t]
\centering
\includegraphics[width=0.32\textwidth]{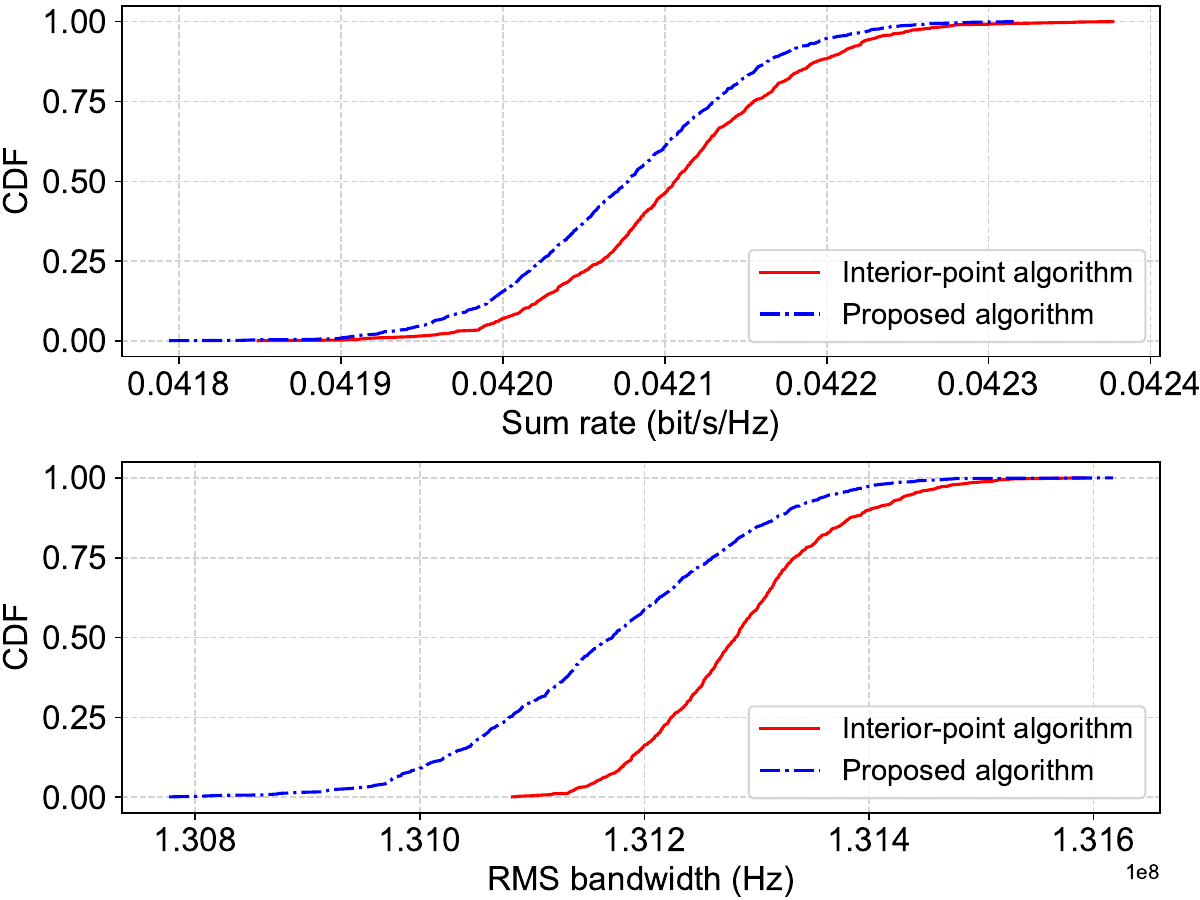}
\caption{Spectral efficiencies of communications and sensing}
\label{fig:exhausted_proposed_comparison}
\end{figure}

\subsubsection{Tradeoff Performance}

To maximize RMS bandwidth, power allocation typically prioritizes frequency edges, a strategy that often conflicts with communication optimization goals. Therefore, a trade-off exists between optimizing these two metrics. We evaluate this trade-off by jointly optimizing sum rate and RMS bandwidth. As illustrated in Fig.~\ref{fig:tradeoff}, the weighting factor $\lambda$ controls the balance between these two metrics. When $\lambda = 0$, the optimization exclusively targets sum rate maximization, achieving the highest sum rate performance of over 0.045 bps/Hz. However, this comes at the expense of RMS bandwidth, which decreases to its lowest value, approximately 124 MHz. Conversely, when $\lambda = 1$, the system prioritizes RMS bandwidth optimization, resulting in the highest RMS bandwidth performance of over 136 MHz, albeit with a degradation in sum rate. Furthermore, as $\lambda$ varies from 0 to 1, the RMS bandwidth consistently increases, whereas the sum rate correspondingly decreases, clearly demonstrating the inherent trade-off between these two performance metrics.

\subsubsection{Comparison with single-stage approach}

We compare the optimization performance of our proposed low-complexity two-stage algorithm with a typical single-stage approach. The common interior-point method serves as our baseline for comparison. A Monte Carlo simulation consisting of $10^3$ runs is conducted, with fast fading channels randomly generated for each run while maintaining constant slow fading conditions. The weighting factor $\lambda$ is set to 0.5. Fig.~\ref{fig:exhausted_proposed_comparison} illustrates the CDF results. From the results, we observe that the interior-point approach slightly outperforms our proposed algorithm. However, the interior-point method demands significantly more computational resources and processing time. The performance gap between the two approaches is minimal, demonstrating that our proposed algorithm achieves near-optimal performance with substantially lower complexity.

\subsection{Experimental Results}

\begin{figure}[!t]
   \centering
   \subfigure[Experiment devices.]
   {
       \includegraphics[width=0.45\columnwidth]{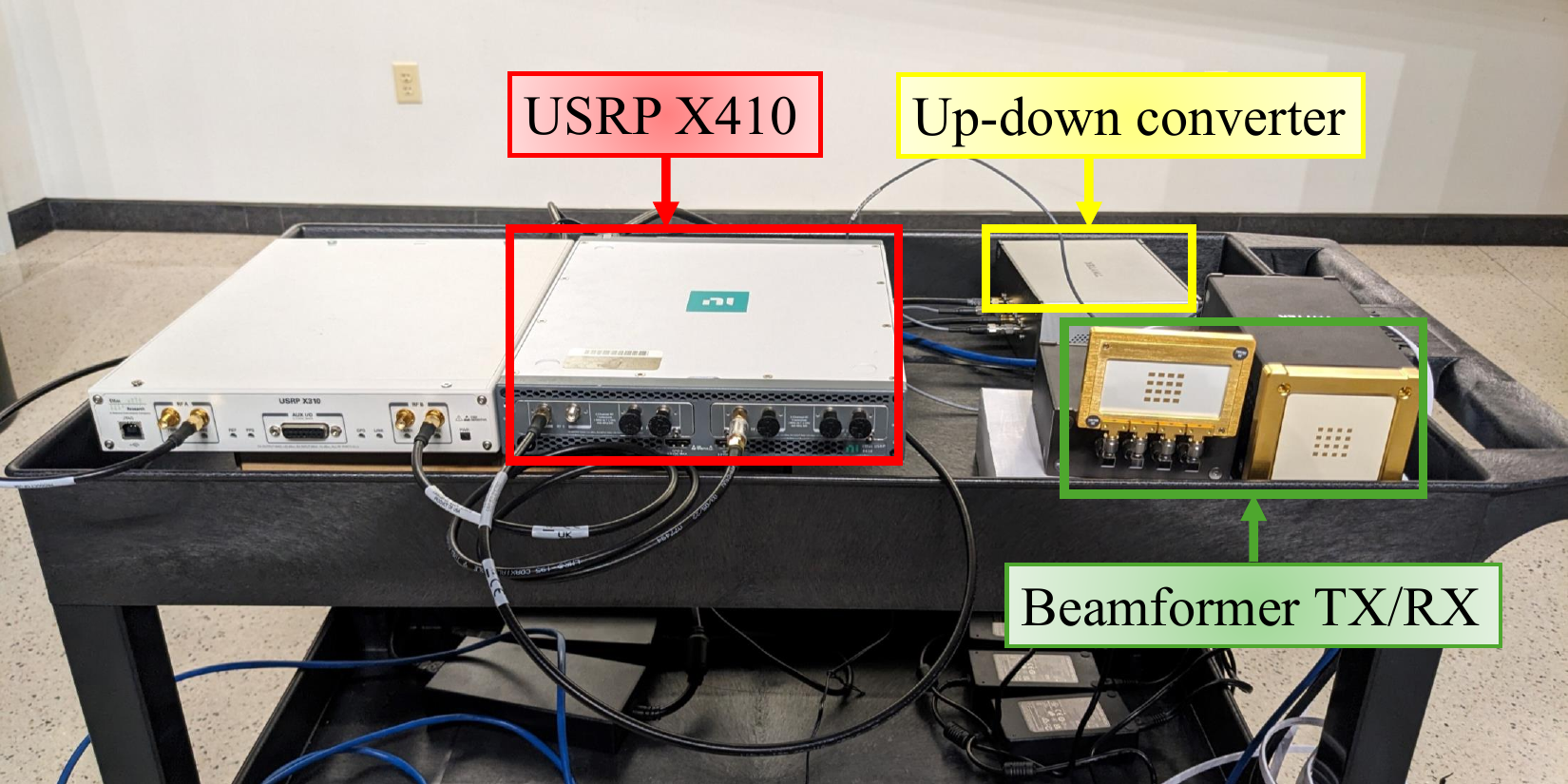}
       \label{fig:devices}
   }
   \subfigure[Experimental setup.]
   {
      \includegraphics[width=0.45\columnwidth]{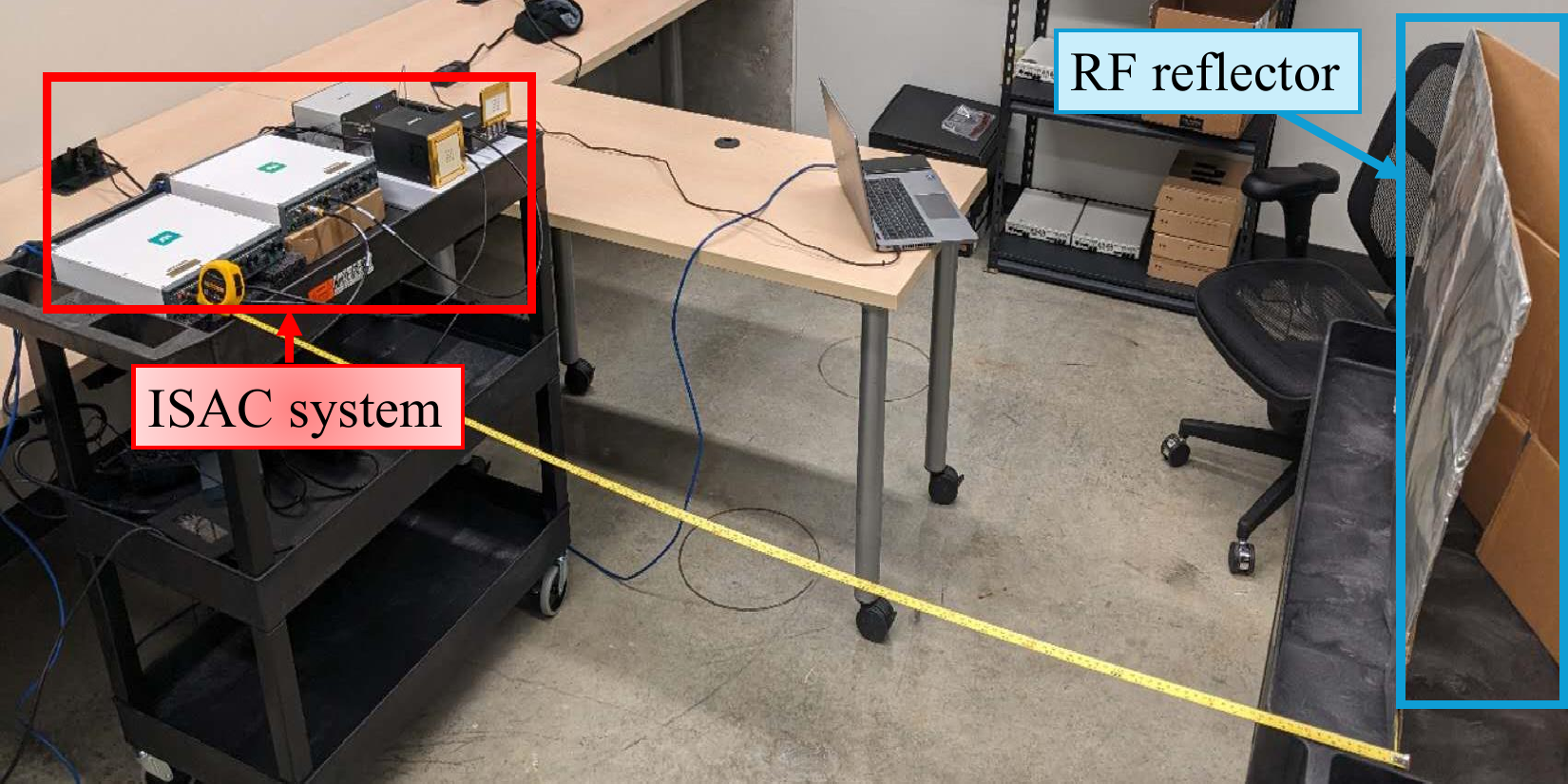}
      \label{fig:experimental_setup}
   }
   \caption{Hardware and experimental setup.}
   \label{fig:setup}
\end{figure}

\begin{figure}[!t]
\centering
\includegraphics[width=0.31\textwidth]{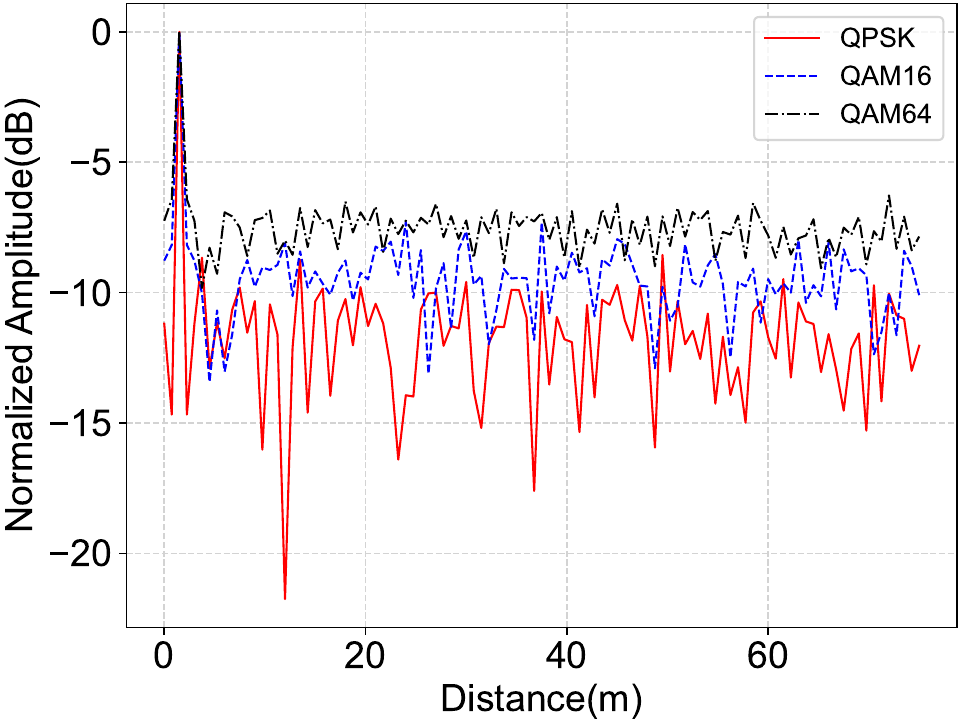}
\caption{Correlation results of signals with different modulation orders.}
\label{fig:QPSK_QAM_corr_experiment}
\end{figure}

The experimental setup, depicted in Fig.~\ref{fig:devices}, utilizes the USRP X410 to generate and capture baseband OFDM signals. An up-down converter is employed to transition the signals between the baseband and the radiofrequency band at 28 GHz. The front-end transceiver comprises separate beamformers for transmission (TX) and reception (RX). Additionally, an RF reflector is positioned at a specified distance from the transceiver, as illustrated in Fig.~\ref{fig:experimental_setup}. 
In Fig.~\ref{fig:QPSK_QAM_corr_experiment}, we visually present the periodic correlation results for QPSK, 16QAM, and 64QAM. It is clear that QPSK exhibits the lowest sidelobe levels, followed by 16QAM, whereas 64QAM shows the most pronounced sidelobes.

\section{Conclusion}\label{sec:conclusions}
In this paper, we studied wideband ISAC waveform synthesis under OFDM with constant modulus phase signaling (for example, QPSK) and transmit PSD shaping for sensing. This decoupled design reveals an OFDM and OTFS duality: flattening the PSD across frequency in OFDM or along the delay axis in OTFS lowers integrated sidelobe levels and sharpens the ambiguity mainlobe. Simulations and an SDR prototype show that a simple PSD variance constraint keeps the spectrum flat, reduces sidelobes versus 16 QAM and 64 QAM, and narrows the mainlobe. To balance sensing and communications in frequency selective channels, we proposed a variance constrained, water filling like allocator that provides a tunable tradeoff between rate and sidelobes and offers a practical way to enforce flatness. Overall, the results confirm robust sidelobe suppression under PSD shaping and support OFDM with QPSK and PSD shaping as a simple and effective wideband ISAC baseline.

\begin{appendices}
    \section{Proof of Theorem \ref{thm:MSE}}\label{appdx:proof}
\begin{proof}
For notational simplicity, we omit the time index in the radar pulses. The Fisher information $I$ for the parameter $\tau$ is given by
\begin{small}
\begin{eqnarray}
I&=&E\left[\left|\frac{d}{d\tau}\log p\left(Y(jw)|\tau\right)\right|^2\right]\nonumber\\
&=&E\left[\left|\frac{d}{d\tau}\left[-\frac{1}{N_0W}\int_{w_c-\frac{W}{2}}^{w_c+\frac{W}{2}} \right.\right.\right.\nonumber\\
&&\left.\left.\left.\left|Y(jw)-Ae^{-jw\tau}X(jw)\right|^2 dw\right]\right|^2\right]\nonumber\\
&=&\frac{1}{N_0^2W^2}E\left[\left|\int_{w_c-\frac{W}{2}}^{w_c+\frac{W}{2}}(Y(jw)-Ae^{-jw\tau}X(jw))\right.\right.\nonumber\\
&\times& \left.\left.\left(jAwe^{-jw\tau}X(jw)\right)^*dw\right|^2\right]\nonumber\\
&=&\frac{1}{N_0^2W^2}E\left[\left|\int_{w_c-\frac{W}{2}}^{w_c+\frac{W}{2}} N(jw)\left(jAwe^{-jw\tau}X(jw)\right)^*dw\right|^2\right]\nonumber\\
&=&\frac{A^2}{N_0W}\int_{w_c-\frac{W}{2}}^{w_c+\frac{W}{2}} w^2|X(jw)|^2dw.
\end{eqnarray}
\end{small}

The conclusion is obtained by the Cramer-Rao bound $MSE\geq \frac{1}{I}$ for unbiased estimators.
\end{proof}
\end{appendices}

\bibliographystyle{IEEEtran}
\bibliography{main}

\end{document}